\documentclass[letterpaper, 10pt,journal,final,oneside]{IEEEtran}
\IEEEoverridecommandlockouts
\ifCLASSINFOpdf
\else
\fi
\usepackage{booktabs} % For formal tables
\usepackage[noadjust]{cite}
\usepackage{amsmath}
\usepackage{amssymb}
\usepackage{amsfonts}
\usepackage{float}
\usepackage{graphicx}
\usepackage[font=footnotesize,labelfont={it}]{caption}
\usepackage{xcolor}
\usepackage{colortbl}
\usepackage{multirow}
\usepackage{mathtools}
\usepackage{algorithm}
\usepackage{algorithmicx}
\usepackage{algpseudocode}
\usepackage{tabto}
\usepackage{hhline}
\usepackage{enumitem}
\usepackage{pifont}
\usepackage[none]{hyphenat}
\makeatletter
\renewcommand\footnoterule{%
  \kern-3\p@
  \hrule\@width 0.5\columnwidth
  \kern2.6\p@}
  \makeatother

\begin{document}
\sloppy
%\setlength\parskip{1pt}
%\linenumbers
\title{MacLeR: \underline{Mac}hine \underline{Le}arning-based \underline{R}un-Time Hardware Trojan Detection in Resource-Constrained IoT Edge Devices}
% -------------- Authors ---------------------------------
%
\author{
    Faiq~Khalid,~\IEEEmembership{Student Member,~IEEE,}
        Syed Rafay Hasan, \IEEEmembership{Member,~IEEE,}
        Sara Zia,
        Osman Hasan, \IEEEmembership{Senior~Member,~IEEE,}
        Falah Awwad, \IEEEmembership{Senior~Member,~IEEE,}
        and~Muhammad Shafique,~\IEEEmembership{Senior~Member,~IEEE}% <-this % stops a space
        \vspace{-0.4in}
% ------------Affiliations ---------------------------------
\IEEEcompsocitemizethanks{
    \IEEEcompsocthanksitem F. Khalid, is with Technische Universit{\"a}t Wien (TU Wien), Vienna, Austria.\protect\\
    E-mail: faiq.khalid@tuwien.ac.at
    \IEEEcompsocthanksitem S. R. Hasan, is with Tennessee Tech. University, Cookeville, TN, USA.\protect\\
    E-mail: shasan@tntech.edu
    \IEEEcompsocthanksitem S. Zia and O. Hasan are with School of Electrical Engineering and Computer Sciences, National University of Sciences and Technology (NUST), Islamabad, Pakistan.\protect\\
    E-mail: \{szia.msee15seecs, osman.hasan\}@seecs.nust.edu.pk
    \IEEEcompsocthanksitem F. Awwad is with United Arab Emirates University, Al Ain, UAE.\protect\\
    E-mail: f\_awwad@uaeu.ac.ae
    \IEEEcompsocthanksitem M. Shafique is with Division of Engineering, New York University Abu Dhabi (NYU AD), Abu Dhabi, United Arab Emirates, and Technische Universit{\"a}t Wien (TU Wien), Vienna, Austria.\protect\\
    E-mail: muhammad.shafique@tuwien.ac.at,muhammad.shafique@nyu.edu
}% <-this % stops an unwanted space
\thanks{Manuscript received April 18, 2020; revised June 12, 2020; accepted July 6, 2020. This article was presented in the International Conference on Compilers, Architecture, and Synthesis for Embedded Systems 2020 and appears as part of the ESWEEK-TCAD special issue.}
}
\markboth{Accepted at CASES 2020 and appears as part of the ESWEEK-TCAD special issue ( Volume: 39, Issue: 11, Nov. 2020)}%
{F. Khalid \MakeLowercase{\textit{et al.}}: ML-based Run-time HT detection}
\maketitle
\begin{abstract} 
Traditional learning-based approaches for \textit{run-time} Hardware Trojan detection require complex and expensive on-chip data acquisition frameworks, and thus incur high area and power overhead. To address these challenges, we propose to leverage the power correlation between the executing instructions of a microprocessor to establish a machine learning-based run-time Hardware Trojan (HT) detection framework, called MacLeR. To reduce the overhead of data acquisition, we propose a single power-port current acquisition block using current sensors in time-division multiplexing, which increases accuracy while incurring reduced area overhead. We have implemented a practical solution by analyzing multiple HT benchmarks inserted in the RTL of a system-on-chip (SoC) consisting of four LEON3 processors integrated with other IPs like vga\_lcd, RSA, AES, Ethernet, and memory controllers.
Our experimental results show that compared to state-of-the-art HT detection techniques, MacLeR achieves 10\% better HT detection accuracy (i.e., 96.256\%) while incurring a 7x reduction in area and power overhead (i.e.,  0.025\% of the area of the SoC and $<0.07\%$ of the power of the SoC). In addition, we also analyze the impact of process variation and aging on the extracted power profiles and the HT detection accuracy of MacLeR. Our analysis shows that variations in fine-grained power profiles due to the HTs are significantly higher compared to the variations in fine-grained power profiles caused by the process variations (PV) and aging effects. Moreover, our analysis demonstrates that, on average, the HT detection accuracy drop in MacLeR is less than 1\% and 9\% when considering only PV and PV with worst-case aging, respectively, which is $\approx$10x less than in the case of the state-of-the-art ML-based HT detection technique. \vspace{-15pt} 
\end{abstract}

% \begin{IEEEkeywords}
% Machine Learning, Hardware Trojans, Power Profiling, Microprocessor, LEON3, Hardware Security
% \end{IEEEkeywords}
%====================================================================
%====================================================================
\section{Introduction}\label{Introduction}
Globalization in an integrated circuit (IC) design-process has exponentially increased the trend to outsource fabrication, which makes the IC designs vulnerable to security threats like Hardware Trojans (HT)~\cite{tehranipoor2010survey,huang2020survey,ratasich2019roadmap,shafique2018intelligent}. 
These HTs can change the system functionality (i.e., the functional Trojans), reduce reliability, increase the likelihood of system failure, modify physical parameters~\cite{li2016survey}\cite{abbassi2019trojanzero} like power consumption, accelerate aging factor~\cite{mossa2017hardware}\cite{mossa2017self}, or contribute to information leakage via side channels (i.e., the parametric Trojans). These consequences can have severe and long-lasting effects on the credibility of hardware, hence, making it imperative to develop efficient HT detection techniques. 

State-of-the-art HT detection techniques utilize side-channel parameters, i.e., timing~\cite{nandhini2018delay,cui2018hardware,babu2019wire}, power~\cite{lodhi2017power}\cite{lodhi2016self}, current, communication~\cite{khalid2018runtime,khalid2020simcom} or electromagnetic signals~\cite{soll2014based}\cite{he2017hardware}, based on the \textit{golden signatures} to detect an anomalous behavior~\cite{lodhi2014hardware}. However, \textit{in the case of third-party-IP based designs, it is nearly impossible to extract the golden signatures because IPs can already be un-trusted}. To address this issue, various IP analysis-based approaches~\cite{zhang2015veritrust,ngo2015hardware,chen2017hardware} have been proposed, but they inherently pose the following limitations:

\begin{enumerate}[leftmargin=*]
    \item A limited access to the IPs and measurement inaccuracies can compromise the accuracy of the golden signatures. 
    \item Reverse engineering-based techniques are costly, and the existing sensors-based techniques \textit{cannot encompass all the possible input conditions} because of the inherent quantization loss of analog-to-digital conversion (ADC).
\end{enumerate} 

To address the above-mentioned limitations, different Machine Learning (ML)-based techniques~\cite{lodhi2017power}\cite{elnaggar2018machine,jap2016supervised,kulkarni2016adaptive,lodhi2016self} have been proposed that train the ML models for communication patterns or power profiles. However, these techniques either possess a large overhead of ML computations or a large overhead of run-time data acquisition, both of which would be infeasible in resource-constrained edge devices, especially under environmental and process variations. Therefore, these issues raise a key research question: \textit{how to enable a lightweight ML-based HT detection technique, and consequently, what is the associated run-time data acquisition overhead and the sensitivity to the process variations?}
\vspace{-0.1in}
\subsection{Motivational Case Study and Key Observations} \label{Motivation}
To address the above question, we propose an alternative approach that exploits the interaction between the trusted and un-trusted IPs in an SoC to extract the corresponding anomalous power profile that can be leveraged to design a low-overhead ML-based HT detection technique. In the context of IoT edge/embedded devices with a shared power distribution network for multiple cores~\cite{flamand2018gap}, \textit{we postulate a hypothesis, ``the activity in an un-trusted IP, interacting with the trusted IP, can have a detectable impact on the power consumption of the trusted IP''.} 
\begin{figure*}[!t]
	\centering
	\includegraphics[width=1\linewidth]{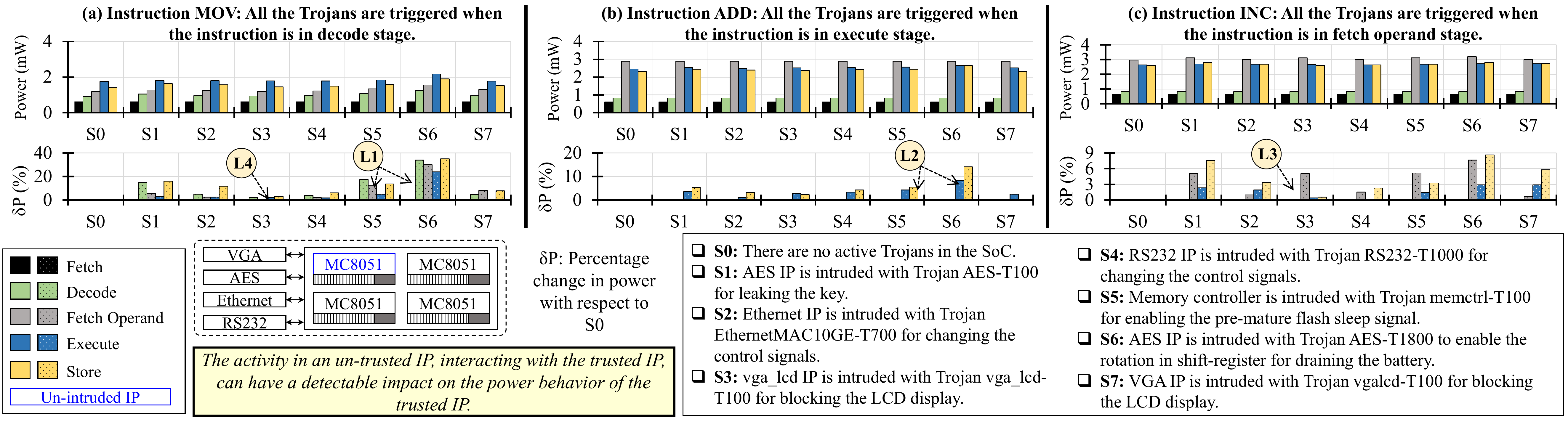} 
	\caption{\textit{Effects of Trust-Hub HT benchmarks (i.e., AES-T100, AES-T800, vga\_lcd-T100, RS232-T1000, memctrl-T100, and ethernetMAC10GE-T700) on the power consumption w.r.t. different pipeline stages of an SoC consisting of four MC8051, one vga\_lcd, one AES, one Ethernet, one RS232, four memory controllers IPs for different instructions, i.e., MOV, ADD, and INC. Note: these results are generated by synthesizing the MC8051 (RTL in Verilog) using the Xilinx toolchain for the Spartan 3(xc3s-1500) FPGA, as we consider resource-constrained IoT edge devices. $\delta P= \frac{|P_{S0}|-|P_{Sx}|}{|P_{S0}|}$, where $P_{S0}$ and $P_{Sx}$ represent the power consumption in scenario S0 and scenario Sx $(x \in \{1, 2, 3, 4, 5, 6, 7\})$, respectively. The workloads used for MC8051 are 32-bit encrypted multiplication, subtraction, addition, and division. The inputs are encrypted using AES, and the results are displayed on the screen using VGA as well as transmitted using Ethernet and RS232 IPs.}}\vspace{-5mm}
	\label{fig:motivation}
\end{figure*}

To corroborate this hypothesis, we perform a proof-of-concept case study (see Fig.~\ref{fig:motivation}) using four MC8051 IPs that are connected with multiple IPs (i.e., vga\_lcd, AES, Ethernet, RS232, memory controllers) in an SoC, and exchange the data with each other to execute a particular set of workloads. Among these MC8051 IPs, at least one MC8051 IP is considered trustworthy, while any of the other IPs can be intruded with (open-source) Trust-Hub HT benchmarks~\cite{trust-HUB} like AES-T100, AES-T800, vga\_lcd-T100, RS232-T1000, memctrl-T100, and ethernetMAC10GE-T700. In these experiments, we monitor the power consumption of the trusted MC8051 to identify the change in power consumption of each instruction in different pipeline components. 

In our experimental results given in Fig.~\ref{fig:motivation}, the top-row shows fine-grained power profiles of the trusted MC8051 microcontroller while executing the instructions (i.e., MOV, ADD, INC) for eight different test scenarios. Each scenario is associated with the activation of a single HT except in the scenario \textit{S0} (i.e., there is no active HT in any IP). The bottom-row shows the change in power consumption ($\delta P$) in different scenarios w.r.t. the power consumption in scenario \textit{S0}. \textbf{From this analysis, we make the following observations:}

\begin{enumerate}[leftmargin=*]
    \item The $\delta P$ in scenario \textit{S6} is a lot higher than the $\delta P$ in other scenarios (see Labels L1 and L2). The reason behind this is that in \textit{S6}, the HT activity (i.e., constant information leakage) is higher than other scenarios. Similarly, some of the scenarios show smaller $\delta P$ values. However, $\delta P$ depends upon instruction. For example, in scenario \textit{S3}, the value of $\delta P$ is $< 2\%$ for the MOV instruction (see Label L4), but for the INC instruction $\delta P$ is $\approx 5\%$. 
    
    \item The change in power consumption $\delta P$ also varies in different pipeline stages. For example, in all instructions and all scenarios, the value $\delta P$ in the Fetch pipeline stage is negligible. However, this value can be higher in unknown HTs. The value of $\delta P$ in the Decode pipeline stage for MOV is higher (see labels L1 and), but the value of $\delta P$ in the Decode pipeline stage for ADD is negligible. \textit{Therefore, for complete coverage of the change in power consumption ($\delta P$) due to an HT, we need to consider the individual power profiles of each pipeline stage.}    
    
    \item In all the scenarios, the values of $\delta P$ are large enough and can be detected\footnote{Note: The detectable change in power consumption depends upon the calibration of the current sensors and the corresponding analog-to-digital converter. For example, some state-of-the-art run-time power analyses cannot detect less than 5\% change in power consumption~\cite{khalidbehavior}.} using state-of-the-art run-time power analysis~\cite{khalidbehavior}. For example, in all the scenarios, the values of $\delta P$ vary from 1\% to 35\%.
    
\end{enumerate}
In summary, these observations strongly indicate that the extracted fine-grained power profiles can be leveraged for run-time ML-based HT detection.

\subsection{Associated Research Challenges} 
The classification of the above-discussed power profiles and monitoring them during run-time, design time, or even during testing leads to the following research challenges:

\begin{enumerate}[leftmargin=*]
    \item In a contemporary SoC, for resource-constrained IoT edge devices, fine-grained power analysis for an n-stage pipeline is not straightforward as it involves a lot of dependencies. This raises a key question about how to extract the distinguishing power profiles at the granularity of different pipeline stages with a minimum overhead?
    
    \item How to exploit these diverse fine-grained power profiles of different instructions (or instruction types) to develop a lightweight ML-based run-time HT detection technique while keeping the complexity and area overhead minimal?

    \item Would the fine-grained power-analysis still be useful to accurately detect HTs at run time under the process variations and aging effects? 

\end{enumerate}
\subsection{Novel Contributions and Concept Overview}
\begin{figure}[!t]
	\centering
	\includegraphics[width=1\linewidth]{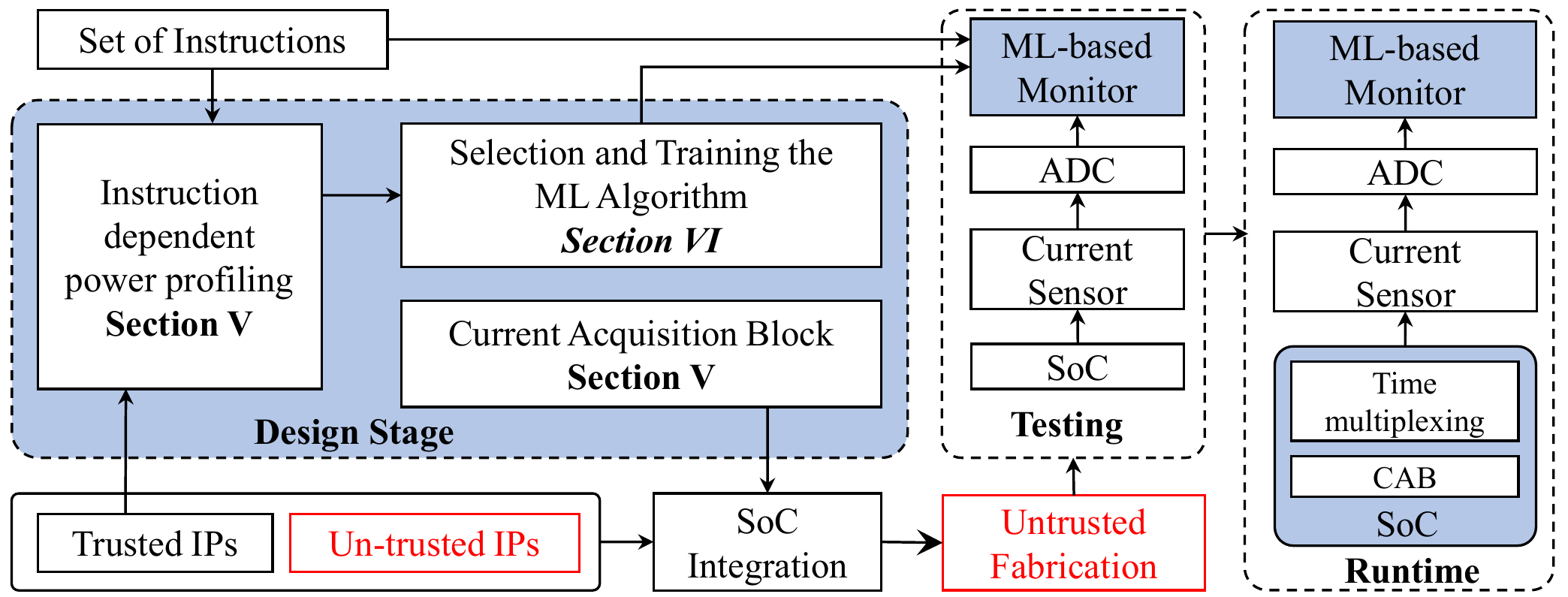} 
	\caption{\textit{Design, test, and run-time flow of our methodology for ML-based HT detection technique (MacLeR). Highlighted boxes represent the novel contributions. The evaluation/testing is done on a LEON3-based SoC.}}
	\label{fig:NC}
	\vspace{-3pt}
\end{figure}

To address the above research questions, we propose a novel methodology, called MacLeR, to design an ML-based run-time HT detection technique that exploits the fine-grained power profiling of the microprocessor (see Fig.~\ref{fig:NC}). Towards this, MacLeR employs the following analysis and methods 
\begin{enumerate} [leftmargin=*]
	\item To obtain the fine-grained power profiles, we propose to measure the individual power of each pipeline stage w.r.t. a particular instruction (see \textbf{Section~\ref{sec:FGPP}}). The reason for choosing this method is that the impact of HTs on fine-grained power profiles is relatively more noticeable as compared to the overall power.
	\item To reduce the complexity and detection time, we propose an off-chip monitor that collects analog power profiles and converts them into the digital domain (see \textbf{Section~\ref{sec:DSE}}). These power profiles are then used first for training an ML model (in our case it is a lightweight multi-layer perceptron (MLP)) at design time, and afterwards at run time for detecting HTs. The reason for choosing an MLP is because it requires fewer computations and is typically faster than other complex ML algorithms. 
	\item Extracting the fine-gained power profiles of a microprocessor during run time requires multiple power ports. Therefore, to reduce the number of power ports, we propose a single power-port current acquisition block (SP-CAB) and accordingly measure the current in a time-division multiplexing manner (see \textbf{Section~\ref{sec:FGPP}}). 
	\item To study the robustness of MacLeR (i.e., drop in HT detection accuracy), we perform a sensitivity analysis under the process variation by performing the Monte-Carlo simulation using the PV models from TSMC 65nm technology (see \textbf{Section~\ref{sec:PV}}).  
	\item To study the robustness of MacLeR (i.e., drop in HT detection accuracy), we also perform a sensitivity analysis under aging effects with and without different aging polices, i.e., Fast-Core-Age-First and balanced-aging profile (see \textbf{Section~\ref{sec:aging}}).
	\item To illustrate the scalability and generalizability of the MacLeR, we evaluated MacLeR for SoC with un-trusted microprocessor IPs and non-processor SoC (see \textbf{ Section~\ref{sec:scalability}}).
\end{enumerate}

\textbf{Hardware Design:} To evaluate our MacLeR framework with the above methods, we developed a LEON3-based SoC consisting of four LEON3 processor IPs integrated with one vga\_lcd IP, one RS232 IP, one Ethernet IP, four memory controller IPs, one basic-RSA and one AES IP for multiple instructions using different ML-algorithms (see Section~\ref{sec:Leon3}).

\textbf{Key Results Compared to the state-of-the-art ML-based HT detection Technique:} We analyzed our MacLeR on the above-mentioned SoC for multiple trust-Hub HT benchmarks and compared it with the state-of-the-art technique presented in~\cite{lodhi2017power}. Key results of these experiments are:

\begin{enumerate}[leftmargin=*]
    \item With only a single port power measuring block, our\textit{ MacLeR, using an MLP with two hidden layers and eight neurons per layer,} achieves HT detection accuracy of 96.256\% that is $\approx$10\% more than the maximum accuracy achieved by the technique presented in~\cite{lodhi2017power}.
    \item The area and power overheads of MacLeR are $\approx$7x less than the area, and power overheads of the technique presented in~\cite{lodhi2017power}. MacLeR requires $\approx7.01\mu m^2$ ($\approx 0.15\%$ of the area of the implemented SoC) and $\approx15\mu W$ ($<1\%$ of the power of the implemented SoC), respectively, when synthesized for 65nm technology using the Cadence Genus.
\end{enumerate} 

\textbf{Sensitivity to the Process Variations and Aging:} The power consumption-based HT detection techniques can be sensitive to process variations (PV) and aging. Therefore, we also analyze the impact of PV on MacLeR by performing Monte-Carlo simulations using the given PV models from TSMC 65nm technology. The aging variations are based on the model presented in literature~\cite{gnad2015hayat,tiwari2008facelift,rehman2014dtune}, i.e., change in operating frequency after Year-1, Year-2, Year-5, and Year-10. We evaluated MacLeR with no aging mitigation and for two aging policies, i.e., Fast-Core-Age-First and balanced-aging profile. Key results of this sensitivity analysis are:  
\begin{enumerate}[leftmargin=*]
    \item In the microprocessor, the power variations due to HT are significantly higher compared to the PV-induced power variations or power variations due to the aging effects. 
    
    \item The average drop in the HT detection accuracy of MacLeR is less than 1\% and 9\% when considering only PV and PV with worst-case aging, respectively, which is $\approx$10x less than in the case of the state-of-the-art ML-based HT detection technique. 
\end{enumerate}
\section {Related Work}
Typically, HT detection techniques employ \textit{delay}~\cite{nandhini2018delay,cui2018hardware,babu2019wire}, \textit{power consumption}~\cite{lodhi2017power,lodhi2016self,lodhi2014hardware} and \textit{operating frequency}~\cite{hou2018chip}\cite{hou2018r2d2} signature-based analysis. However, most of these techniques are payload-specific, i.e., they can only detect HTs  during the design or testing phases, and require golden circuits. However, the intruder may hide HTs by \textit{exploiting the aging behavior}~\cite{mossa2017hardware}\cite{mossa2017self} or \textit{the switching activity}~\cite{abbassi2019trojanzero} of the chip. Detecting such HTs during the testing phase is very challenging, and the undetected HTs may get activated once the chip is in use. Run-time approaches, on the other hand, can monitor an IC for its entire operational lifetime, providing an important last-line of defense. Therefore, specialized techniques have been developed to detect HTs with specific payload at runtime, i.e., confidentiality~\cite{veeranna2017trust}, integrity~\cite{malekpour2017trojanguard} and availability~\cite{zhao2015applying} attacks. \textit{The main drawback of such run-time techniques is that they incur large area and power overhead~\cite{bhunia2013protection} and require precise calibration to cater to environmental changes and process variations.}

\begin{table}[!t]
	\caption{\textit{Comparison with state-of-the-art ML-based run-time HT detection (SVM: support Vector Machine, RE: Reverse Engineering, TMR: Triple Modular Redundancy, DT: Decision Tree,``n'': Number of components involve in a pipeline operation, C: Confidentiality, I: Integrity, A: Availability, P: Power, D: Delay)}}
	\label{tab:literature}
	\resizebox{1\linewidth}{!}{
    		\begin{tabular}{|c|c|c|c|c|c|c|c|}
            \hline
            \multirow{2}{*}{\textbf{Techniques}} & \multicolumn{3}{c|}{\textbf{Payloads}} & \multicolumn{2}{c|}{\textbf{Parameters}} & \multirow{2}{*}{\begin{tabular}[c]{@{}c@{}}ML\\ Tools\end{tabular}} & \multirow{2}{*}{\begin{tabular}[c]{@{}c@{}}On-Chip Overhead\end{tabular}} \\ \cline{2-6}
             & \textbf{C} & \textbf{I} & \textbf{A} & \textbf{P} & \textbf{D} &  &  \\ \hline
            \cite{bao2014application,bao2016reverse} &  &  & \checkmark & \checkmark & \checkmark & SVM & \begin{tabular}[c]{@{}c@{}}SVM and RE overhead\end{tabular} \\ \hline
            \cite{kulkarni2016adaptive} & \checkmark & \checkmark &  &  & \checkmark & SVM & \begin{tabular}[c]{@{}c@{}}SVM Model and TMR\end{tabular} \\ \hline
            \cite{lodhi2017power,khalidbehavior} &  & \checkmark & \checkmark & \checkmark &  & \begin{tabular}[c]{@{}c@{}}DT \end{tabular} & \begin{tabular}[c]{@{}c@{}}``n'' Power-ports and ADCs\end{tabular}\\ \hline
            MacLeR & \checkmark & \checkmark & \checkmark & \checkmark &  & \begin{tabular}[c]{@{}c@{}}MLP\end{tabular} & \begin{tabular}[c]{@{}c@{}}One power-port, ``n'' current \\ sensors,  and Time Multiplexer\end{tabular} \\ \hline
        \end{tabular}
    }
\end{table}
 
To address the above-mentioned limitations, ML-based techniques have emerged as a promising solution to detect the possibility of anomalies at run time, while exploiting the datasets obtained during the measurement phase for training~\cite{elnaggar2018machine}. \textit{One of the major challenges in such techniques is generating the parametric or behavioral profile to train the ML tool.} Recently, a support vector machine (SVM) has been used to classify the intruded and un-intruded parametric behavior~\cite{jap2016supervised}. However, modeling and acquiring the dynamic behavior of SoC using SVM is computationally costly, e.g., it requires $m\times p$ multiplications and requires $p\times wordsize$, where $m$ and $p$ are the size of data and number of features, respectively. To cater to this problem, Kulkarni et al.~\cite{kulkarni2016adaptive} proposed to use other supervised ML online algorithms, i.e., k-NN and Modified Balanced Winnow (MBW) algorithm. However, these approaches assume that IP modules are not intruded and therefore, they are only applicable to availability attacks. The work in ~\cite{lodhi2017power} proposed an on-chip power-based technique, which can detect HTs having direct or indirect effects on the power consumption of a microcontroller~\cite{lodhi2017power}. \textit{However, this HT detection technique requires a large number of power-ports to extract the power behavior for ML training and ML inference, and is therefore infeasible to be deployed in real-world embedded systems}. 
Table \ref{tab:literature} summarizes state-of-the-art ML-based HT detection approaches w.r.t. their features, as well as the orientation of our proposed MacLeR framework highlighting the key differences.\normalcolor

\section{Threat Model}
\begin{figure}[h]
	\centering
	\includegraphics[width=1\linewidth]{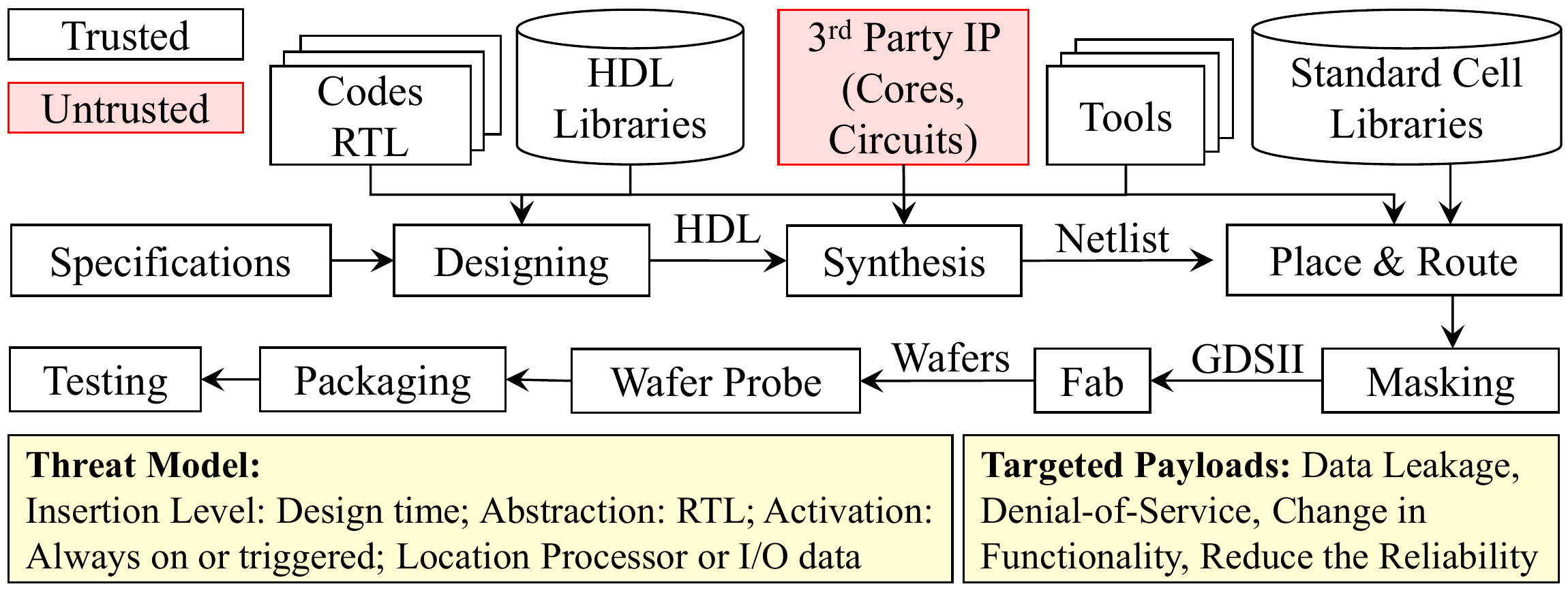} 
	\caption{\textit{Brief overview of the IC manufacturing along with the targeted threat model and the corresponding payloads.}}
	\label{fig:TM}
\end{figure}

We assume that the \textit{third-party IP (3PIP) vendors are not trustworthy}, and therefore, the specification and source code provided by the vendor may contain HTs; see Fig.~\ref{fig:TM}. 

\textit{The hardware designers/architects who integrate different 3PIPs, along with the in-house IPs (if any), to develop an SoC are considered to be the defenders}. Note that among these IPs \textit{at least one IP is trusted.} This paper targets HTs (see Fig.~\ref{fig:TM}), which have a \textit{direct or indirect impact on the shared power network between IPs of a SoC}. Although the multi-core SoCs can have multiple power grids and different voltage islands, the SoCs in battery-operated components for edge devices typically have only one power grid with multiple voltage islands (for instance, GAP-8~\cite{flamand2018gap}), which is shared between different components. Therefore, in this work, \textit{we design a low-power machine-learning-based run-time HT detection methodology for SoCs that have one power grid with multiple voltage islands.}

\section{MacLeR: ML-based Run-time Monitoring}\label{sec:MacLeR}
\begin{figure}[h]
	\centering
	\includegraphics[width=1\linewidth]{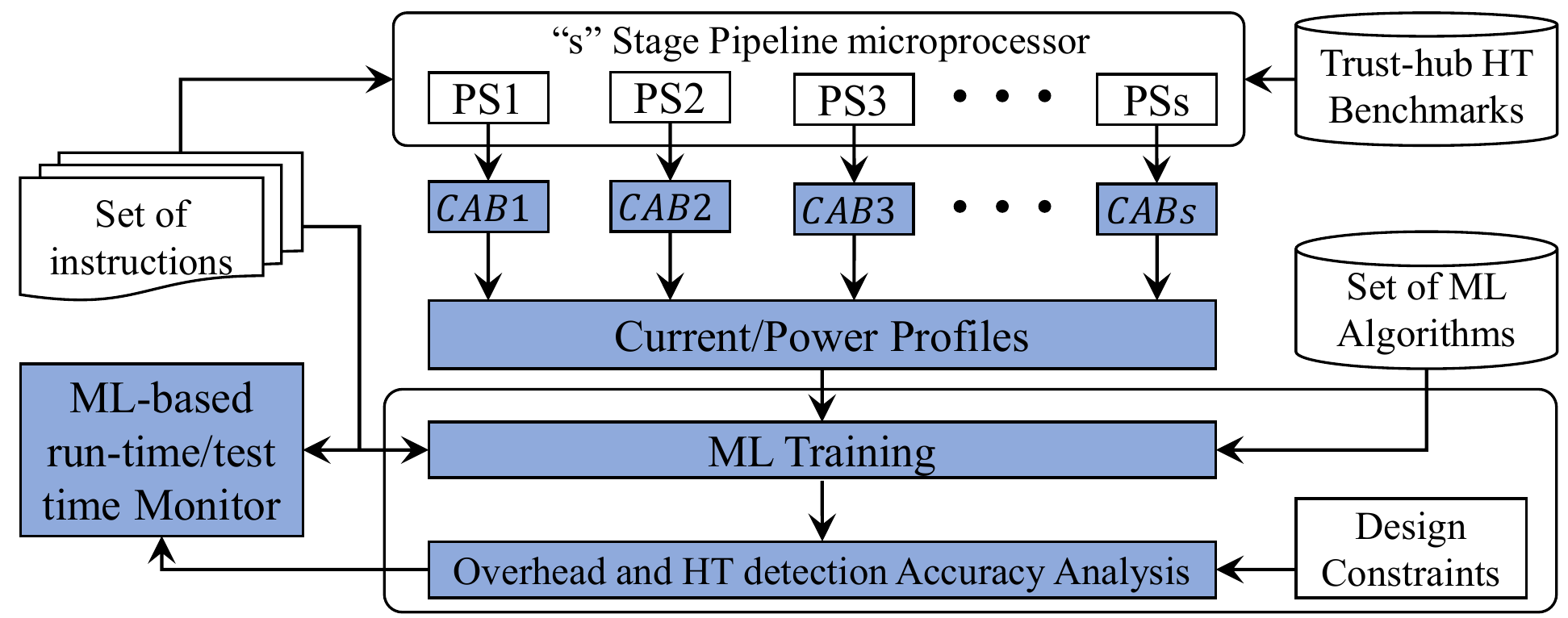} 
	\caption{\textit{MacLeR: ML-based methodology to run-time power monitoring. CABs represent the current acquisition blocks for $s^{th}$ pipeline stages (PSs).}}
	\label{fig:methodology}
\end{figure}

Fig.~\ref{fig:methodology} shows the complete step-by-step flow of McLeR, which consists of the following key steps:
\begin{enumerate}[leftmargin=*]
    \item The main goal of MacLeR is to design a low-power ML-based monitor for HT detection, for which a proper training dataset is required. Towards this, during the design phase, first, MacLeR generates power profiles by measuring the combined power consumption of each component involved in a particular pipeline stage. However, to generate the abnormal power profiles of a microprocessor for MLP training, we use the Trust-Hub HT benchmarks. Note, we use different sets of HT benchmarks for training and testing of MacLeR to avoid any kind of training bias. 
    
    \item During the design phase, it uses the generated power profiles to train different variants of MLPs. Then it performs a design space exploration (DSE) w.r.t. the detection accuracy and the associated overhead of the MLP, and it chooses the most appropriate MLP, which is used for run-time HT detection. 
    
    \item During the run time, MacLeR measures the power consumption of each component involved in a particular pipeline stage using multiple\textit{ current mirrors}\footnote{Current mirror is an analog circuit that copies the current of an active device using diode connected CMOS transistor. Typically, these circuits are used to sense the current of an active device.}, and then it collects the combined power using a single pMOS transistor. Then these \textit{power values are collected in a time-division multiplexing manner} and used by the trained MLP (which is the best-one from the Step-2 of DSE) to detect HTs.

\end{enumerate}
\vspace{-4mm}
\section{Fine-Grained Power Profiling}\label{sec:FGPP}
During design time, fine-grained power profiles are obtained by measuring the instruction-dependent power of each component associated with a pipeline stage. However, power acquisition during the run-time poses a research challenge about \textit{how to acquire the fine-grained power profiles with a minimum area overhead}?
\begin{figure}[!t]
	\centering
	\includegraphics[width=\linewidth]{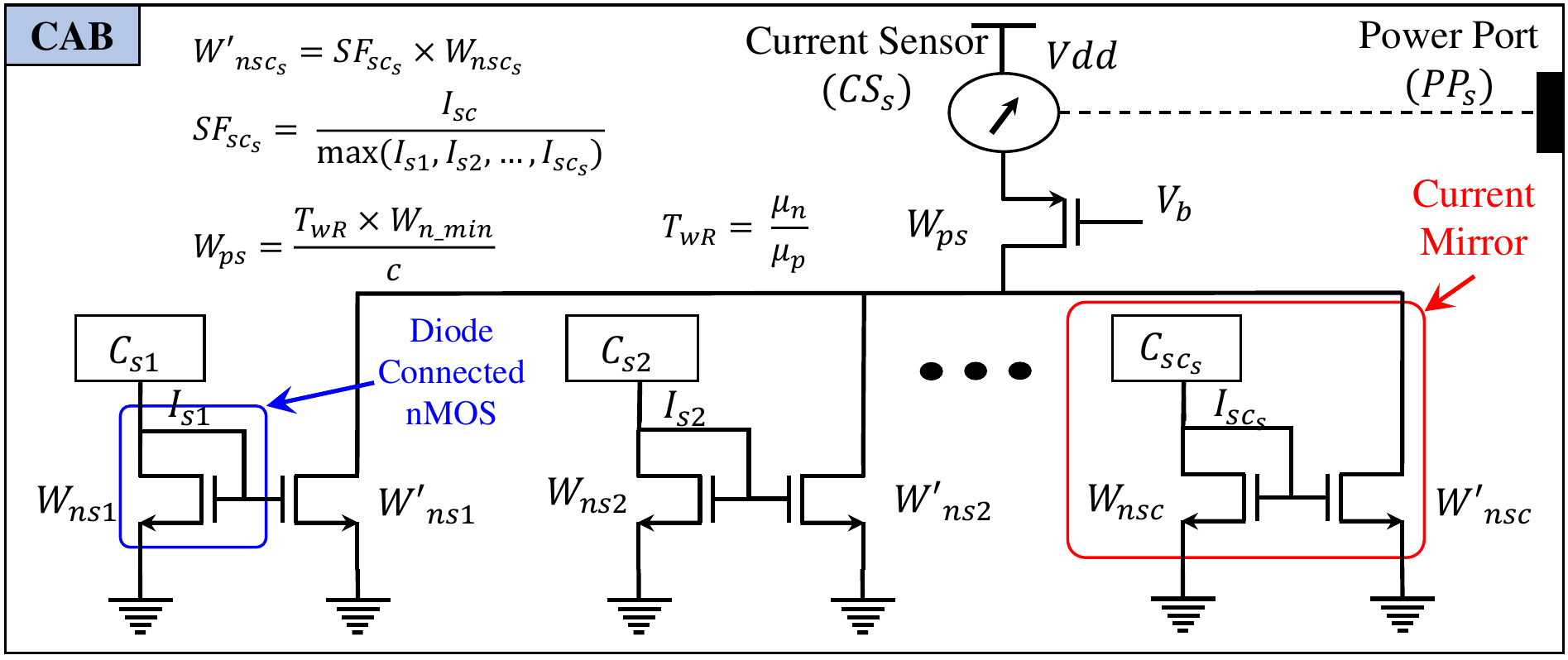} 
	\caption{\textit{The proposed single power-port current acquisition block (CAB).}}
	\label{fig:CAB}
\end{figure}
\begin{figure}[!t]
	\centering
	\includegraphics[width=1\linewidth]{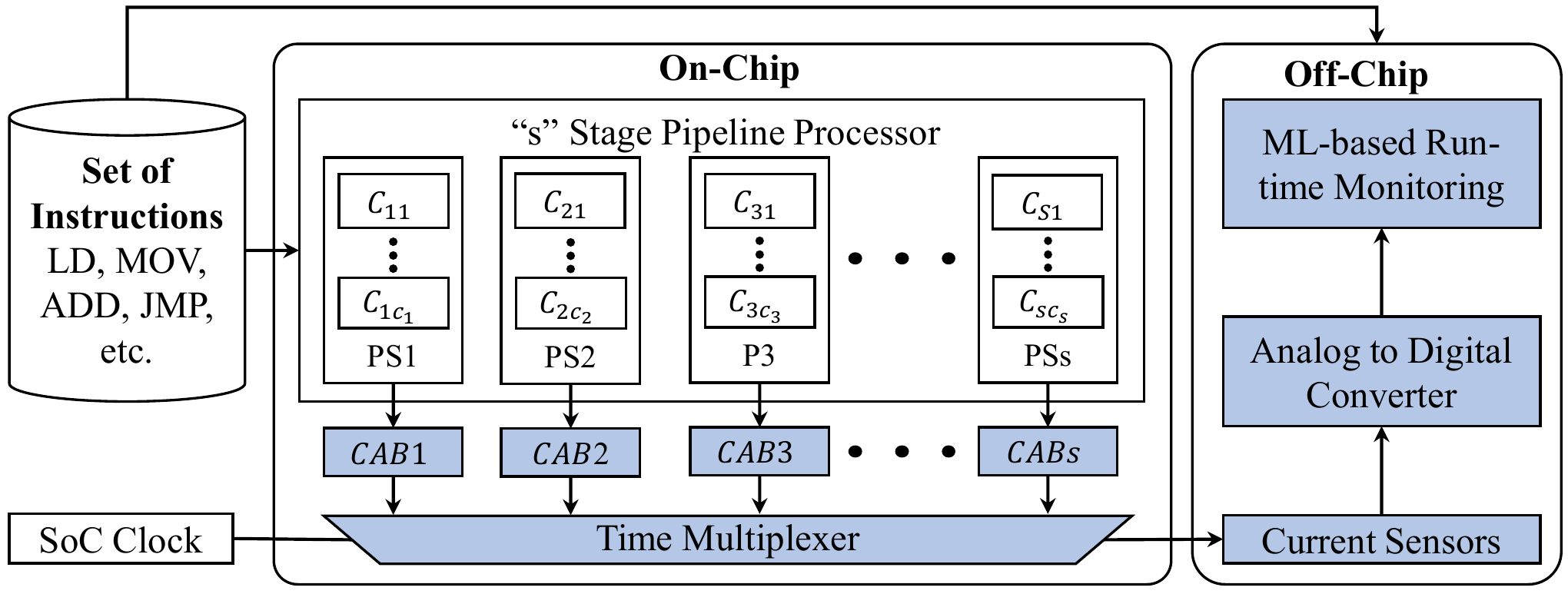} 
	\caption{\textit{Single power-port based hardware implementation of MacLeR with on-chip and off-chip hardware components.}}
	\label{fig:MacLeR_HW}
\end{figure}

\begin{figure*}
	\centering
	\includegraphics[width=1\linewidth]{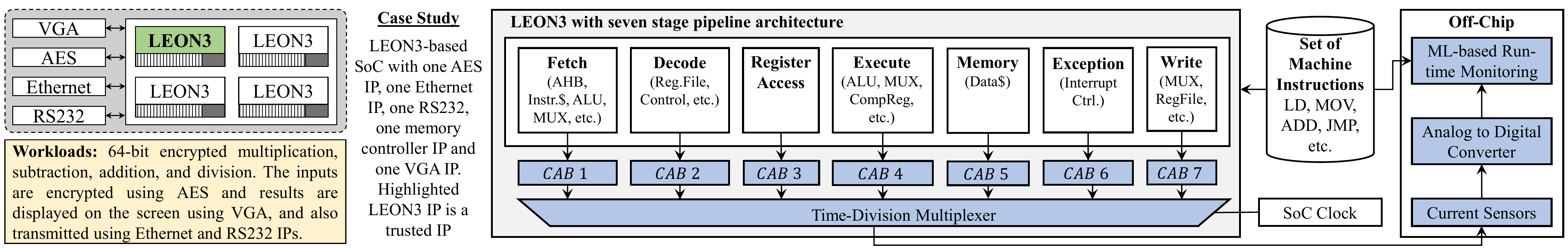} 
	\caption{\textit{Hardware implementation of the MacLeR monitoring framework for LEON3 with a seven-stage pipeline architecture.}}
	\label{fig:HW_LEON}
	\vspace{-12pt}
\end{figure*}
For this, we propose to use \textit{multiple current mirrors} for measuring the current from each component and collect it using a single pMOS transistor-based collector, as shown in Fig~\ref{fig:CAB}. Since the current measuring accuracy is dependent on the sizes of the transistors, therefore, we computed these sizes using the following set of equations:

\begin{equation}\footnotesize
	W'_{nsc_s} = SF_{nsc_s} \times W_{nsc_s} 
	\label{eq:3} 
\end{equation}

\begin{equation}\footnotesize
	SF_{nsc_s} = \frac{I_{nsc_s}}{max(I_{ns1}, I_{ns2},..., I_{nsc_s})}  
	\label{eq:4} 
\end{equation}

Where $W_{nsc_s}$ and $W'_{nsc_s}$ are the widths of nMOS in the parent branch and corresponding mirror branches for $c_s th$ component of the $sth$ pipeline stage, respectively. In some cases, the width requirement exceeds the maximum allowed width. to address this, we introduce the scaling factor $SF_{nsc_s}$ that normalizes the current and respective transistor width, to keep the width requirement within the limit, as shown in Equation~\ref{eq:4}. Similarly, using the model of~\cite{razavi2002design}, we computed the width of a single pMOS for a given component 'C', using the following equation:
	\begin{equation} \footnotesize
	    W_{ps} = \frac{T_{wR} \times W_{n\_min}}{c}\ ;\ define\ \ T_{wR} = \frac{\mu_n}{\mu_p} 
	\label{eq:5} 
	\end{equation}

The single current sensor-based power-ports are used to measure the current of different modules involved in each pipeline stage. Typically, the number of power-ports required to cover all the modules is equal to the number of components involved in the operations of pipeline stages~\cite{lodhi2017power}\cite{khalidbehavior}, e.g., in n-stage pipeline architecture, the number of power-ports is $N_p = C_1 + C_2 + ... + C_n$, where $C_1, C_2, ... , C_n$ are the number of components involved in pipeline operations of $1^{st}$, $2^{nd}$, ... , $n^{th}$ pipeline stage, respectively. Having multiple power ports in an IC is very expansive for the IC packaging. Therefore, to acquire the complete power profile of the microprocessor via a single power-port, we propose to acquire the data using time multiplexing (which measures the current of each SP-CAB after every clock cycle), as shown in Fig.~\ref{fig:MacLeR_HW}. Note, \textit{MacLeR extracts and uses the fine-grained power profiles of a trusted microprocessor in an SoC. Therefore, it does not require any golden circuits of un-trusted IPs}.

\section{Training and Selection of An Efficient MLP Model}\label{sec:DSE}
After acquiring the power profiles of the microprocessor, MacLeR chooses an appropriate ML algorithm based on the required HT detection accuracy and design constraints. Towards this end, we propose an iterative methodology that first trains the multiple configurations of MLPs using the following steps, as shown in Fig.~\ref{fig:methodology}. 
\begin{enumerate}[leftmargin=*]
    \item We start by labeling different power-profiles to differentiate the intruded and un-intruded power profiles. These labelings are, in turn, used to train and validate the ML models. Note, during the design time, the abnormal power profiles are obtained using the trust-hub HT benchmarks.
    	
    \item Next, we categorize these power profiles w.r.t. the functional and behavioral similarity to increase the efficiency of the ML models. 
    	
    \item After the categorization, we train the multiple ML models and validate them by applying the testing dataset. 
    
    \item Finally, MacLeR selects the best MLP model based on the maximum HT detection rate, associated overhead, and the given design constraints. 
\end{enumerate}

Note, MacLeR requires the instruction, its category, and power value, irrespective of the pipeline stage, as shown in Fig.~\ref{fig:HW_LEON}. The main reason to choose the fine-grained power profiling at the pipeline stage is that MacLeR can explore multiple power values to expand its search space.

\section{Case Study: Employing MacLer to A LEON3-Based SoC}\label{sec:Leon3}
We illustrate the practicality, utilization, and effectiveness of our MacLer framework by applying it on a LEON3-based SoC, where at least one of the IP is trusted, as shown in Fig.~\ref{fig:HW_LEON}. The main motivation of choosing LEON3 is that it is highly configurable and open-source, and the HT benchmarks that are provided by trusthub.org~\cite{trust-HUB} can easily be integrated into it. The workloads used for LEON3 are 64-bit encrypted multiplication, subtraction, addition, and division. The inputs are encrypted using AES and results are displayed on the screen using VGA and also transmitted using Ethernet and RS232 IPs. Note, the addition, subtraction, and multiplication are used for training the ML monitor, and the division is used at the inference stage of the ML-monitor to detect the HTs, and therefore, avoiding the biasing in the testing phase.  

\subsection{LEON3: Power Profiling} \label{sec:L_PP}
\begin{figure}[!t]
	\centering
	\includegraphics[width=0.98\linewidth]{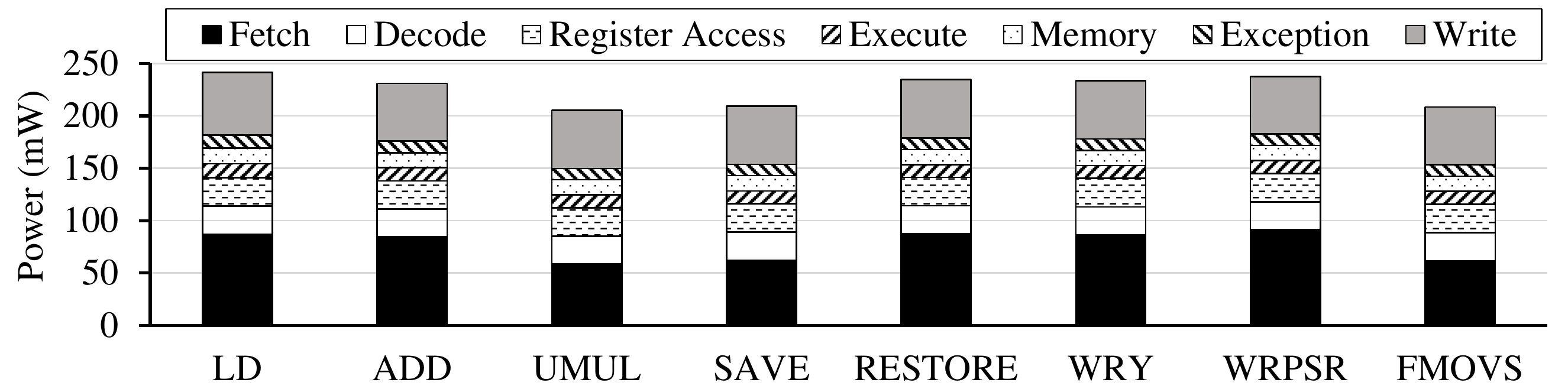}
	\caption{\textit{Power Profiles of un-intruded LEON3 for different instructions}}
	\label{fig:power-profile}
\end{figure}

For obtaining the power profile, we synthesized the LEON3 processor using Cadence Genus (Encounter) tool with the TSMC 65nm library. The power of each module involved in the pipeline stage is calculated separately for each instruction. For example, Fig.~\ref{fig:power-profile} shows the power consumption in each pipeline stage for a particular instruction, which is extracted by executing one instruction at a time. \textit{It is very challenging and computationally expansive to cover all the possible combinations of instructions. Therefore, to reduce the overheads, we use the instruction categorization of SPARC V8 architecture~\cite{sparc1992sparc},} i.e., the following five categories:
\begin{enumerate}[leftmargin=*]
	\item \textbf{Category 1 (Cat1):} \textit{Load/store instructions} access memory and, use registers and a signed 13-bit immediate value to calculate a 32-bit, byte-aligned memory address, e.g., load (LD), load double (LDD) and load floating-point (LDF).
	\item \textbf{Category 2 (Cat2):} \textit{Arithmetic/logical/shift instructions} perform arithmetic, logical, and shift operations, e.g., subtract (SUB), add (ADD), multiply (ULMUL) and add with carry (ADCC).
	\item \textbf{Category 3 (Cat3):} \textit{Control-transfer instructions} perform PC-relative branches and calls, register-indirect jumps, and conditional traps, e.g., restore (RESTORE), call and link (CALL) and save (SAVE).
	\item \textbf{Category 4 (Cat4):} \textit{Read/Write Register instructions} read and write the contents of software visible state registers and processor registers, e.g., write to processor state register (WRPSR), write to y register (WRY) and read from y register (RDY).
	\item \textbf{Category 5 (Cat5):} \textit{Floating-point operate instructions} perform all floating-point operations, e.g., floating-point move (FMOV), floating-point operate (FCMP) and floating-point add (FADD).	 
\end{enumerate}

Note, we assume all the required data is available in the on-chip memory, and thus, the difference in power profiles during cache hits or misses is not considered in this work. However, this behavior can be captured by exploiting the pipeline stall flags~\cite{schoeberl2018patmos}. 

\begin{figure}[!t]
	\centering
	\includegraphics[width=1\linewidth]{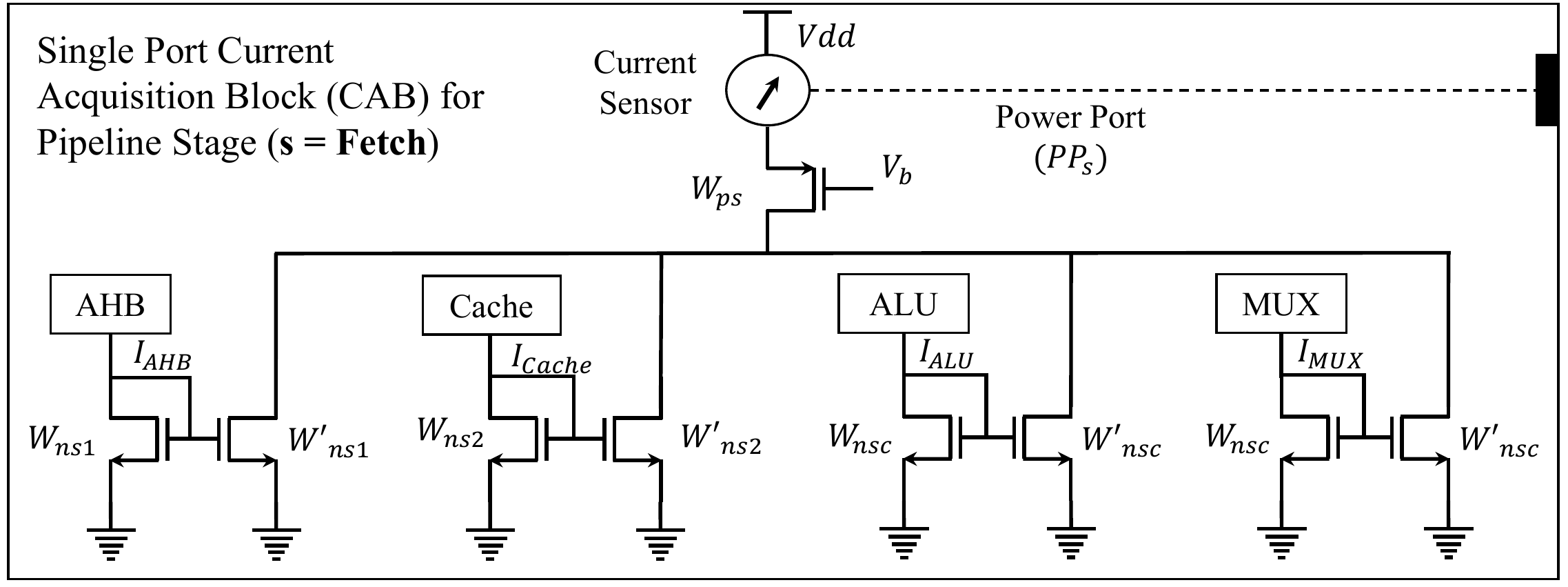} 
	\caption{\textit{Hardware Implementation of the single power-port current acquisition block for pipeline stage \textit{Fetch} of the LEON3 processor.}}
	\label{fig:CAB_LEON}
\end{figure}

\subsection{LEON3: Data Acquisition} \label{sec:L_Dq}
The power consumption of a microprocessor is dependent upon the number of modules involved in the execution of instructions during a particular pipeline stage, e.g., LEON3 has seven different power profiles, i.e., one for each pipeline stage. To model the power behavior, first, we identify all the modules involved in the operation of a particular pipeline stage. Fig.~\ref{fig:HW_LEON} shows that in LEON3, the fetch stage requires instruction cache (I-Cache), AHB bus, adder, and multiplexers (MUX). Decode, register access, execute, memory and exception stages require register file, ALU, data cache (D-cache), and interrupt controller. In LEON3, the fetch stage consists of 4 components, which is the maximum number of components involved as compared to any other pipeline stage. To obtain the power profiles, we implemented the CAB for each pipeline stage of the LEON3, e.g., Fig.~\ref{fig:CAB_LEON} shows the CAB for Fetch stage. To generate the \textit{anomalous power profiles}, we implemented the \textit{trust-hub HT benchmarks}. In our experimental setup, AES-T100, AES-T800, vgalcd-T100, RS232-T1000, memctrl-T100, and ethernetMAC10GE-T700 benchmarks are used to generate the data for \textit{training}, while the data generated using all the HT Trojan benchmarks for RS232, MC8051, AES, Basic RSA, VGA-LCD, memory controller and Ethernet IPs are used for \textit{evaluation}. 

\begin{figure}[!t]
	\centering
	\includegraphics[width=1\linewidth]{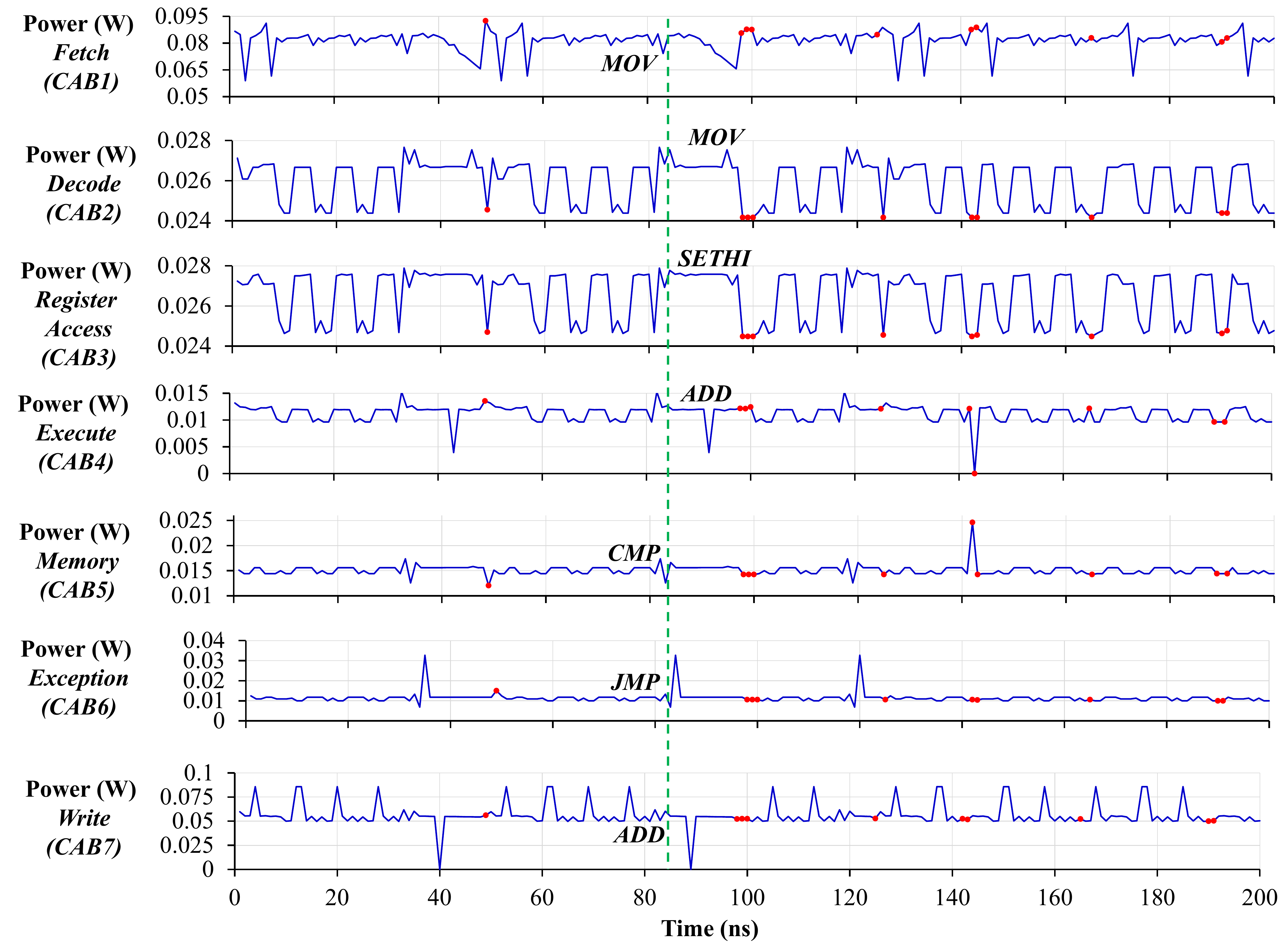} 
	\caption{\textit{Power consumption of seven pipeline stages of the LEON3 microprocessor while executing multiple instructions. The Red dots ({\tiny \color{red}{\ding{108}}}) in the figure represent the activations of MC8051-T200 in the SoC.}}
	\label{fig:power}
\end{figure}

Fig.~\ref{fig:power} shows the power profiles of LEON3 in the presence of AES-T100 in SoC, and the red dots ({\tiny \color{red}{\ding{108}}}) show the triggering of AES-T100. The power value at each time unit, along with an instruction and its category, is extracted to generate the required power profile\footnote{For example, at 1ns the power profile consists of the power of each pipeline stage is ([0.086649W, 0.027123W, 0.027238W, 0.013182W, 0.015111W, 0.012424W, 0.059794W], LD and Cat1).}. These profiles are then used to train the ML model in the next phase.

\begin{figure}[!t]
	\centering
	\includegraphics[width=1\linewidth]{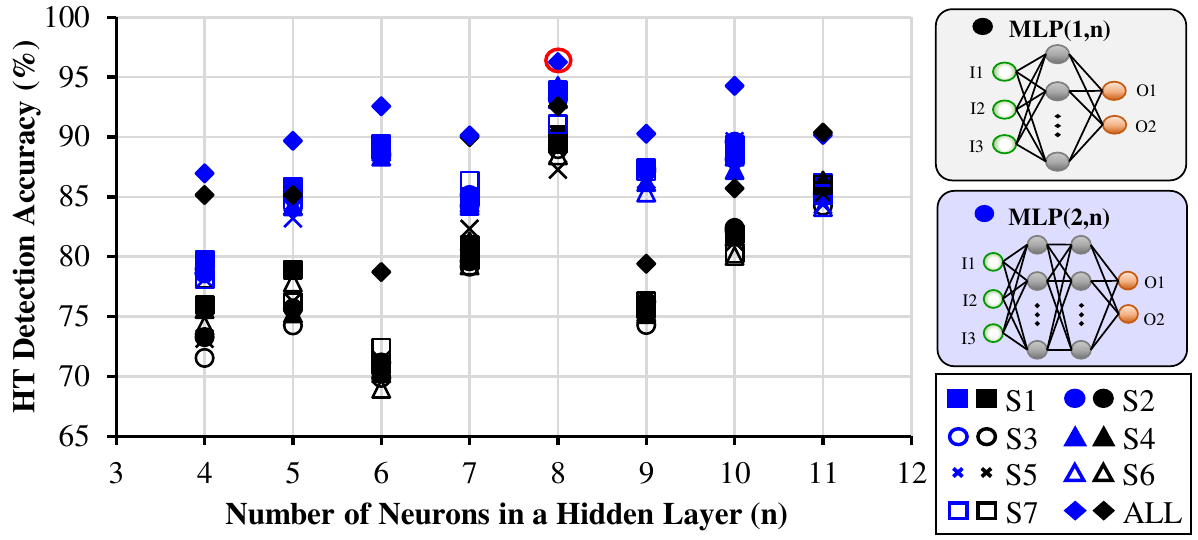} 
	\caption{\textit{Design space exploration of different configurations of implemented MLPs with one or two hidden layers. The blue color represents the MLP with two hidden layers, and the black color represents the MLP with one hidden layer. The inputs for MLP are I1: Power, I2: instruction, and I3: category, and output labels are O1: un-intruded and O2: Intruded.}}
	\label{fig:MLP}
\end{figure}
\begin{figure*}[!t]
	\centering
	\includegraphics[width=1\linewidth]{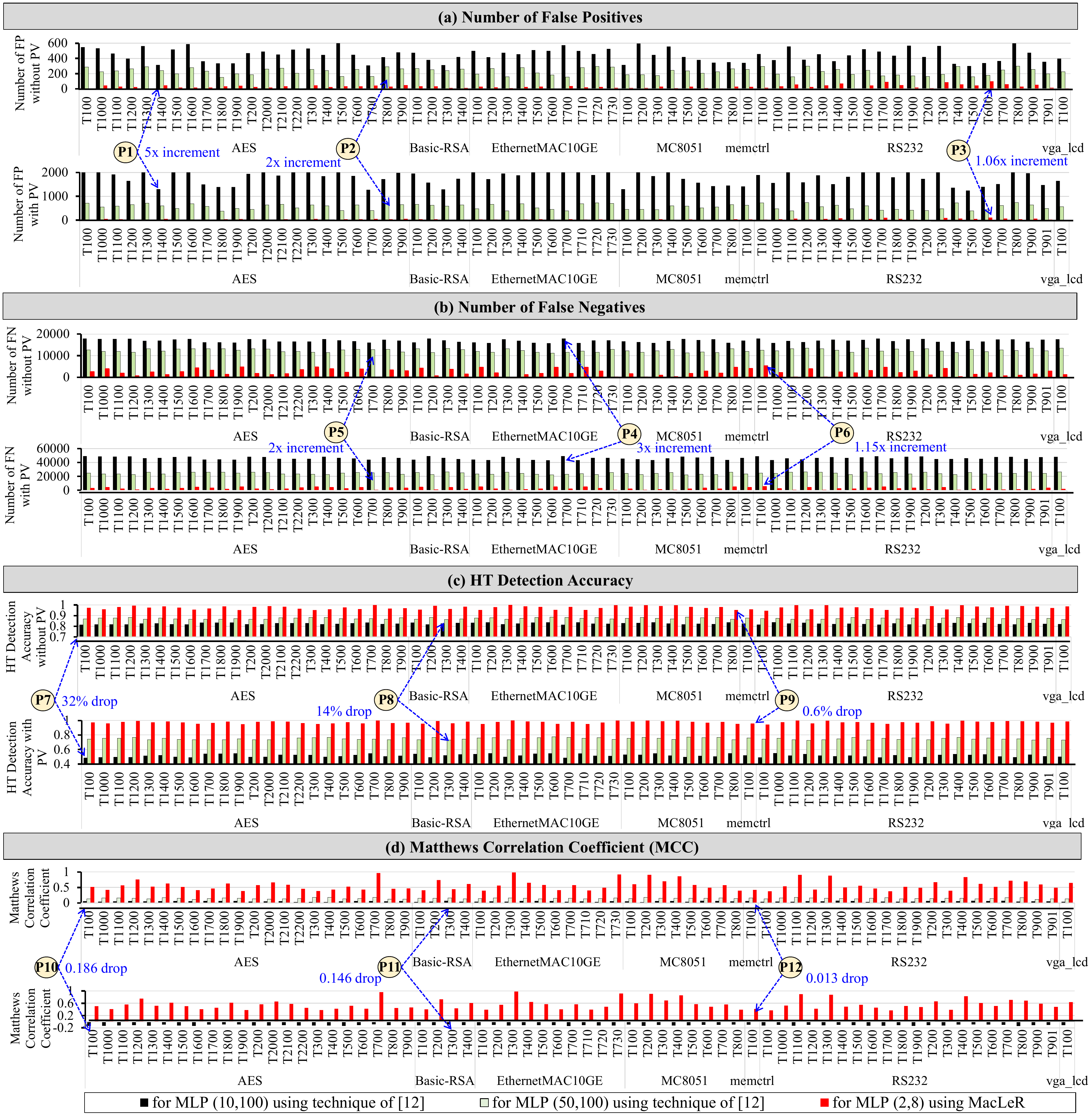} 
	\caption{\textit{False positives, false negatives and HT detection accuracy of the MacLeR and the state-of-the-art run-time ML-based HT detection technique~\cite{lodhi2017power}, in the presence of different HT benchmarks \textbf{with} and \textbf{without} considering the process variations. Note: these analyses are based on the 100,000 classification per HT, where the total number of activations is 1000 out of 100,000. In these experiments, all benchmarks related to MC8051 are implemented in the LEON3 microprocessor and all other HT benchmarks are also configured for the LEON3 microprocessor. In these experiments, the overall accuracy is computed as $Accuracy = \frac{TP+TN}{TP+TN+FP+FN}$, where TP, TN, FP, and FN represent true-positives, true-negatives, false-positives, and false-negatives, respectively. The Mathews Coefficient is computed using the standard formula, $MCC = \frac{(TP\times TN)-(FP\times FN)}{\sqrt{(TP+FP)\times(TP+FN)\times (TN+FP)\times(TN+FN)}}$.}} 
	\vspace{-4mm}
	\label{fig:FP_FN}
\end{figure*}
\subsection{LEON3: Run-time Monitor for HT Detection} \label{sec:L_RM}
After extracting the power profile in the previous phase (see Fig.\ref{fig:power}), we analyze and transform it in such a way that it can be utilized for training and validation. This involves proper labeling of the extracted power behavior to differentiate between \textit{intruded} and \textit{un-intruded} behaviors, and pre-processing to reduce the redundant dataset. After the transformation, we trained different configurations of neural networks (multi-layer perceptrons (MLP) with one and two hidden layers, as shown in Fig.~\ref{fig:MLP}). Afterwards, we analyzed the trade-off between accuracy and computational cost as shown in Fig.~\ref{fig:MLP}. For a comprehensive analysis, we trained each configuration on the abnormal power profiles extracted from the different scenarios (explained in Fig.~\ref{fig:motivation}). For example, MLP (1,8) is trained for anomalous data (generated from the scenarios S1, S2, S3, S4, S5, S6, and S7 individually, and a combination of all scenarios) and the normal data (generated from the scenario S0). This analysis shows that the MLP with two hidden layers and eight neurons in each hidden layer provides maximum HT detection accuracy, i.e., 96.256\% in our case study.   

\textbf{HT Detection Accuracy:} To validate the extracted trained model, we randomly divide the labeled data into $k$-mutually exclusive subsets, where each subset is approximately of the same size, and performs the training and validation $k$ times. In each iteration, one subset is used for validation, and the others are used in training. Thus, each subset is used for an equal number of times for training and once for validation. We also evaluated the trained network for unseen data generated using trust-hub HT benchmarks, as shown in Fig.~\ref{fig:FP_FN}. 

The experimental analysis shows that in the case HT benchmarks for MC8051, the trained MLP with two hidden layers and eight neurons in one layer provides approximately 98\% HT detection accuracy, and with a very small number of false positives and false negatives. However, for other HT benchmarks, MacLeR still provides approximately 90\% HT detection accuracy. \textit{Based on these observations, we conclude that the impact of HTs on shared power network, especially in multi-IP based SoC with at least one trusted microprocessor, can be detected by observing the fine-grained power profiles of the trusted processor.} 

Moreover, for comprehensive analysis, we also compute the Mathews correlation coefficient, as shown in Fig.~\ref{fig:FP_FN}(d). MLP1(10, 100) and MLP2(50,100) using~\cite{lodhi2017power} gives up to 88\% HT detection accuracy. However, the MCC analysis shows that MLPs trained using~\cite{lodhi2017power} are either randomly flipping the binary decision (because MCC is close to zero, see labels P10 and P11) or show the negative correlation. The reason for this is that number of false positives, and the number of false negatives is very large. Therefore, these MLPs cannot be considered as a good binary classifier. On the other hand, MCC values for an MLP trained using MacLeR go up 0.7 (see label P12), which is the property of a very good binary classifier.

Note that MacLeR can sense only one stage at a time; this may affect its HT detection accuracy. However, typically, the activation time (including triggering and payload time) for HT is more than a couple of clock cycles~\cite{trust-HUB}, and the sampling frequency of the MacLeR is equal to the maximum operating frequency of the microprocessor. Therefore, the number of false negatives for MacLeR is very small.
\section{Sensitivity Analysis of MacLeR under Process Variations} \label{sec:PV}
The side-channel parameters (like Power) based-HT detection techniques are vulnerable to environmental changes (e.g., temperature variations and measurement noise) and process variations (PV). To illustrate the variation-tolerance of MacLeR, we perform an analysis using the following steps:

\begin{figure*}[!t]
	\centering 
	\includegraphics[width=1\linewidth]{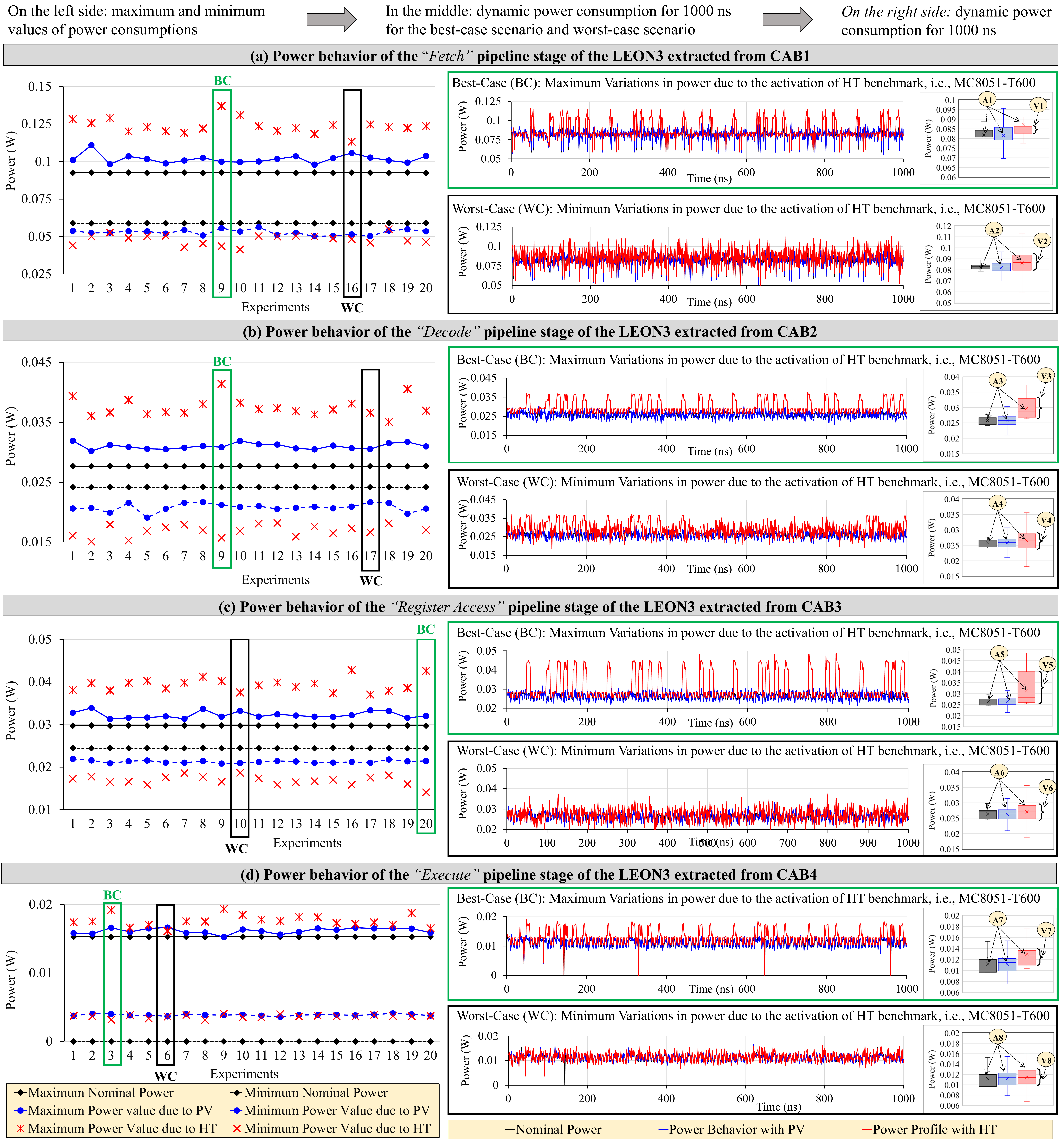} 
	\caption{\textit{The impact of process variations and implemented hardware Trojan benchmark, i.e., MC8051-T600, on the power behavior of \textbf{Fetch, Decode, Register Access}, and \textbf{Execute} pipeline stages of LEON3 microprocessors while executing the multiple instructions. The results presented in black, blue and red colors represent the nominal power behavior, variations in power behavior due to PV, and variations in the power behavior due to HT, respectively. Note, in these analyses, Best-Case (BC) defines the scenarios in which HT is easily detectable, and Worst-Case (WC) defines the scenarios in which HT is hard to detect.}} 
	\vspace{-4mm}
	\label{fig:PV-1}
\end{figure*}
\begin{figure*}[!t]
	\centering 
	\includegraphics[width=1\linewidth]{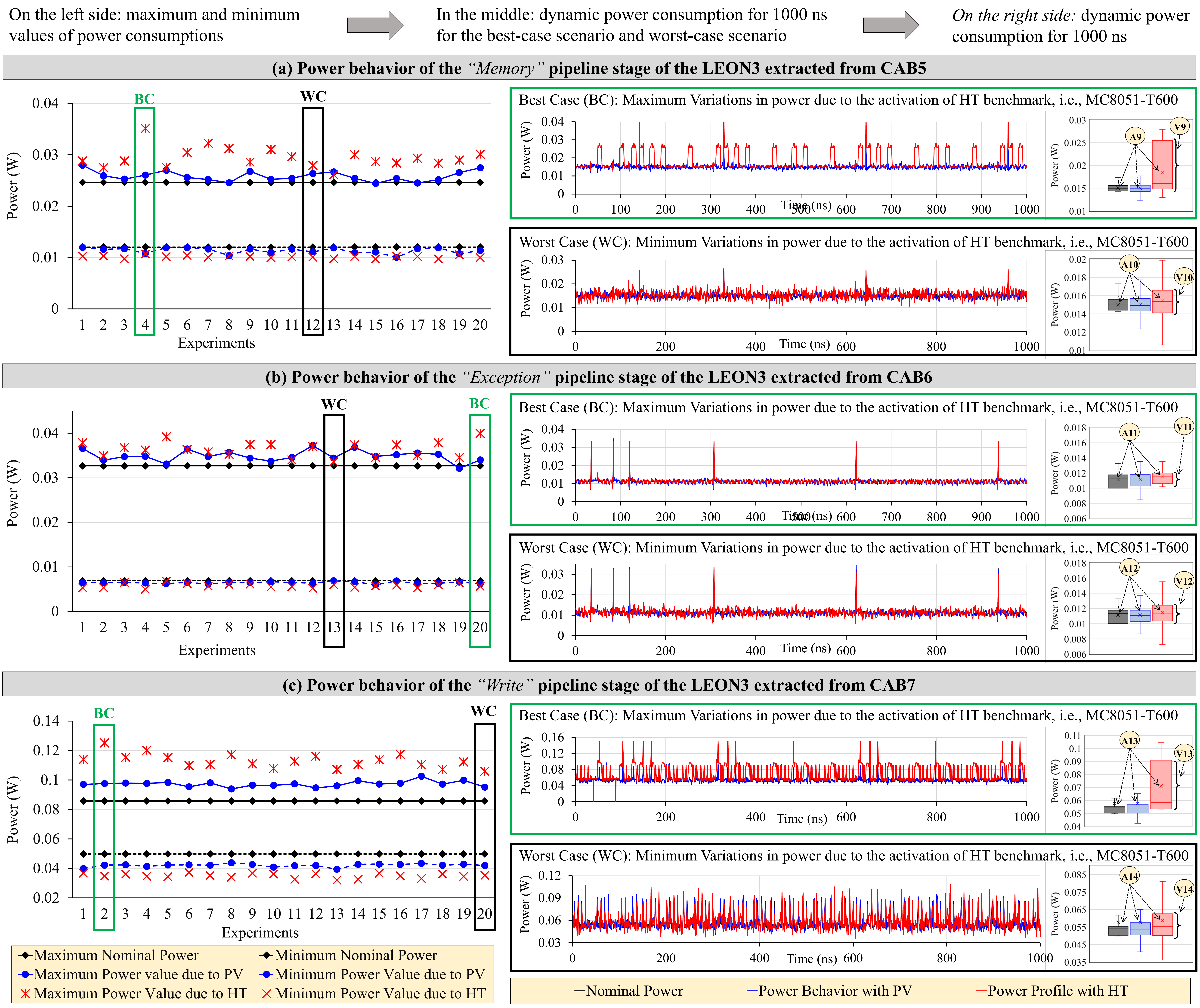} 
	\caption{\textit{The impact of process variations and implemented hardware Trojan benchmark, i.e., MC8051-T600, on the power behavior of \textbf{Memory, Exception}, and \textbf{Write} pipeline stages of LEON3 microprocessors while executing the multiple instructions. The results presented in black, blue and red colors represent the nominal power behavior, variations in power behavior due to PV, and variations in the power behavior due to HT, respectively. Note, in these analyses, Best-Case (BC) defines the scenarios in which HT is easily detectable, and Worst-Case (WC) defines the scenarios in which HT is hard to detect.}} 
	\vspace{-4mm}
	\label{fig:PV-2}
\end{figure*}
\begin{enumerate}[leftmargin=*]
    \item First, we perform the Monte-Carlo analysis using the given PV model from TSMC 65nm technology to generate power profiles with and without HT activation while using multiple workloads, i.e., the addition, subtraction, and multiplication. In this experiment, we perform 100 experiments for each workload with and without each benchmark Trojan.  
    
    \item Afterwards, we use these power profiles to train the MLP that is selected in Section VII-C based on the highest HT detection accuracy (see Fig.\ref{fig:MLP}). Then, the trained MLP is used to analyze the impact of PV on the HT detection accuracy. Note, we use the division as the workload for analyzing the HT detection accuracy of the trained MLP. 
    
\end{enumerate}

\subsection{Impact of PV on Fine-Grained Power Profiles}
The impact of process variations is not uniformly distributed for an SoC. Depending upon the fabrication conditions and variation in the fabrication process, it affects different components with different intensities. Therefore, to analyze the impact of process variation on fine-grained power profiles, we individually analyze the power profiles extracted from the CAB associated with each pipeline stage of the LEON3 microprocessor. For example, Figs.~\ref{fig:PV-1} show the impact of PV on fine-grained power profiles of LEON3 microprocessor with and without HT benchmark, i.e., MC8051-T600\footnote{For these analyses, we performed 100 experiments for each case with multiple HT benchmarks. However, due to limited space, we present analysis for 20 experiments for each case with one HT benchmark, i.e., MC8051-T600.}. From these analyses, we made the following observations:  
\begin{figure}[!t]
	\centering 
	\includegraphics[width=1\linewidth]{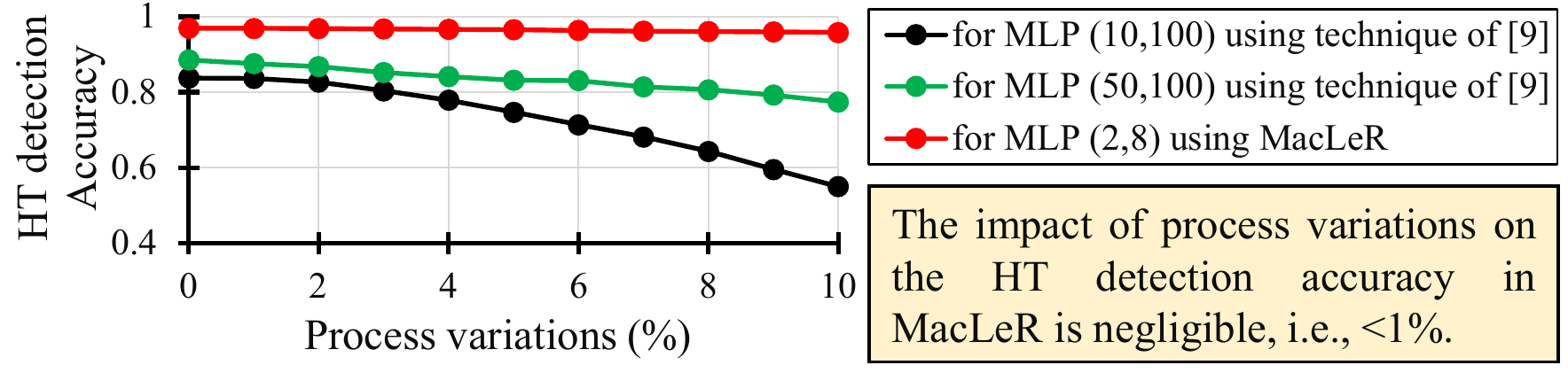} 
	\caption{\textit{Impact of different range of the process variations on the average detection accuracy. Note, this analysis is performed for the MC8051-T600.}} 
	\vspace{-1mm}
	\label{fig:accuracy}
\end{figure}

\begin{figure*}[!t]
	\centering 
	\includegraphics[width=1\linewidth]{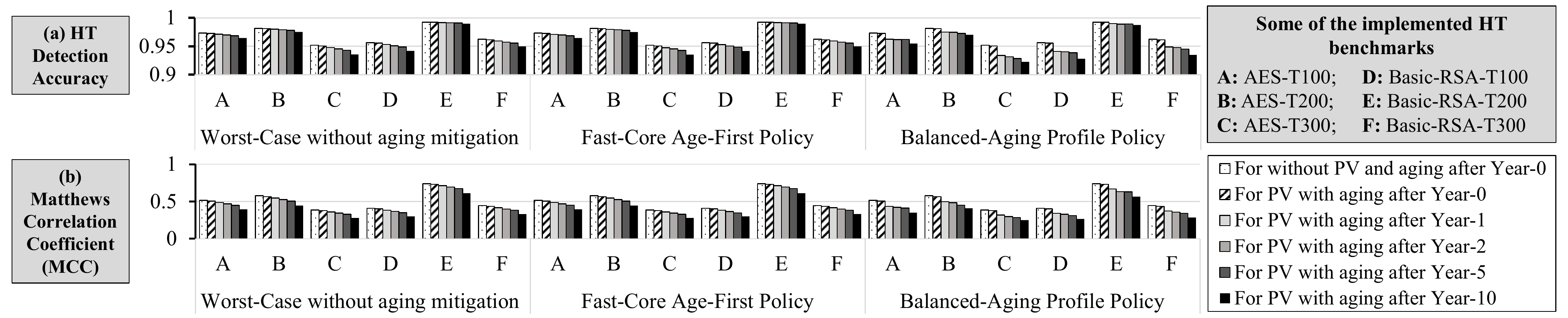} 
	\caption{\textit{Performance analysis of MacLeR for different HT benchmark under different PV and aging variations, i.e., change in operating frequency after Year-1, Year-2, Year-5, and Year-10.}} 
	\vspace{-5mm}
	\label{fig:aging}
\end{figure*}
\begin{enumerate}[leftmargin=*]
    \item The analysis presented in Fig.~\ref{fig:PV-1} (a) shows that in the presence of an active HT (i.e., MC8051-T600) the maximum power consumption of the ``Fetch'' pipeline stage is significantly larger than the PV boundaries\footnote{It is defined as the absolute difference between the maximum ($P_{max}$) and minimum ($P_{min}$) values of power consumption in the presence of PV, $ |B_{PV}| = |P_{max}| - |P_{min}|$.}. For example, the best-case (Experiment 9) analysis shows that the spread of power consumption in the case of MC8051-T600 is smaller (see label V1) but is shifted towards the higher power consumption (see Label A1). This behavior shows the additive nature of MC8051-T600. However, in some cases, the variations are not significant. For example, in the worst-case (Experiment 16) analysis, the variations in power are smaller, and the mean value of power consumption in the presence of MC8051-T600 is within the PV boundaries (see label A2).    
    
    \item Fig.~\ref{fig:PV-1} (b) shows a similar trend for the ``Decode'' pipeline stage. However, variations in power consumption due to  MC8051-T600 are even larger than the variations in power consumption of the ``Fetch'' pipeline. For example, in the best-case (Experiment 9) analysis, the power spread is large (see label V3), and it completely lies outside the PV boundaries.  
    
    \item Similar trends are observed for the other pipeline stages of the LEON3 microprocessor, i.e., ``Register Access'' in Fig.~\ref{fig:PV-1} (c)'', ``Execute'' in Fig.~\ref{fig:PV-1} (d), ``Memory'' in Fig.~\ref{fig:PV-2} (a) and ``Write'' in Fig.~\ref{fig:PV-2} (c). However, the power behavior of the pipeline stage ``Exception’’ is hardly affected by MC8051-T600. For example, the best-case analysis (see labels A11 and V11) and the worst-case analysis (see labels A12 and V12) in Fig.~\ref{fig:PV-2} (b) shows that in most of the experiments, power variations due to MC8051-T600 are within the PV boundaries. The reason behind this is that MC8051-T600 does not get triggered when the instructions are inside the exception pipeline stage. Moreover, in most of the workloads, this pipeline stage remains dormant.      
    
\end{enumerate}
In summary, on average, the HT detection accuracy drops in MLP1(10, 100) and MLP2(50,100) using~\cite{lodhi2017power} are $\approx$ 29\% (83.75\% to 55.002\%) and $\approx$ 11\% (88.56\% to 77.401\%), respectively. However, on average, the HT detection accuracy drop in MacLeR (subjected to PV) is less than 1\% (96.256\% to 95.85\%) compared to the case without PV consideration.

% \color{blue}
% \textit{Based on the above-mentioned analyses, we conclude that in most cases, variations in fine-grained power profiles of the microprocessor due to HT are significant enough (even in the case of abrupt changes like blue peaks in Fig.~\ref{fig:PV-2} (b)) to detect the HT, hence captured by our MacLeR. Moreover, depending upon the applications, the frequency and significance of the pipeline stage can be adjusted to increase the HT detection accuracy.}
% \normalcolor
\subsection{Impact of PV on HT Detection Accuracy}
To analyze the tolerance of MacLeR against PV, we generated the power profiles for different ranges of process variations, i.e, 1\% to 10\%, and analyze the MLP trained using MacLeR, as shown in Fig.~\ref{fig:accuracy}. \textit{This analysis shows that the drop in HT detection accuracy of MLP trained using MacLeR is negligible.} However, the drop in HT detection accuracy of MLPs that are trained using the power profiling technique in~\cite{lodhi2017power} is significant. \textit{The reason behind this is that the stat-of-the-art technique~\cite{lodhi2017power} does not consider the correlation between the instructions with power profiles of microprocessor w.r.t. different pipelines stages.} 

To further illustrate the effectiveness of the MacLeR, we analyze the MLP that is trained using MacLeR for various HT benchmarks from trust-Hub~\cite{trust-HUB} in the presence of 10\% PV. Fig.~\ref{fig:FP_FN} shows the impact of  
10\% PV on the number of false positives, false negatives, and the HT detection accuracy of the trained MLP. From this analysis, we made the following observations: 
\begin{enumerate}[leftmargin=*]
    \item For MLPs that are trained using the power profiling technique in~\cite{lodhi2017power}, in the worst case, the number of false positives is increased by 5x (see label P1 in Fig.~\ref{fig:FP_FN} (a)) and 2x (see label P2 in Fig.~\ref{fig:FP_FN} (a)). However, on average, the number of false positives is increased from 500 (without considering the PV) to 1500 (when considering the PV). On the other hand, the increment in false positives for MacLeR is almost negligible, i.e., 1.06x, see label P3 in Fig.~\ref{fig:FP_FN} (a). A similar trend can be observed for increment in the number of false negatives, i.e., 5x increment, 2x increment, and 1.15x increment as shown by labels P4, P5 and P6 in Fig.~\ref{fig:FP_FN} (b), respectively. 
    
    \item Similar to the analysis, presented in Fig.~\ref{fig:accuracy}, in the worst case, the HT detection accuracy MLPs that are trained using the power profiling technique in~\cite{lodhi2017power} is 32\% (see label P7 in Fig.~\ref{fig:FP_FN} (c)) and 14\% (see label P8 in Fig.~\ref{fig:FP_FN} (c)), respectively. On the other hand, the worst-case drop in the HT detection accuracy for MacLeR is negligible, i.e., 0.6\% as shown by label P9 in Fig.~\ref{fig:FP_FN} (c). In short, in all the cases, on average, the drop in HT detection accuracy is also negligible. 
\end{enumerate}
\begin{figure}[!t]
	\centering 
	\includegraphics[width=1\linewidth]{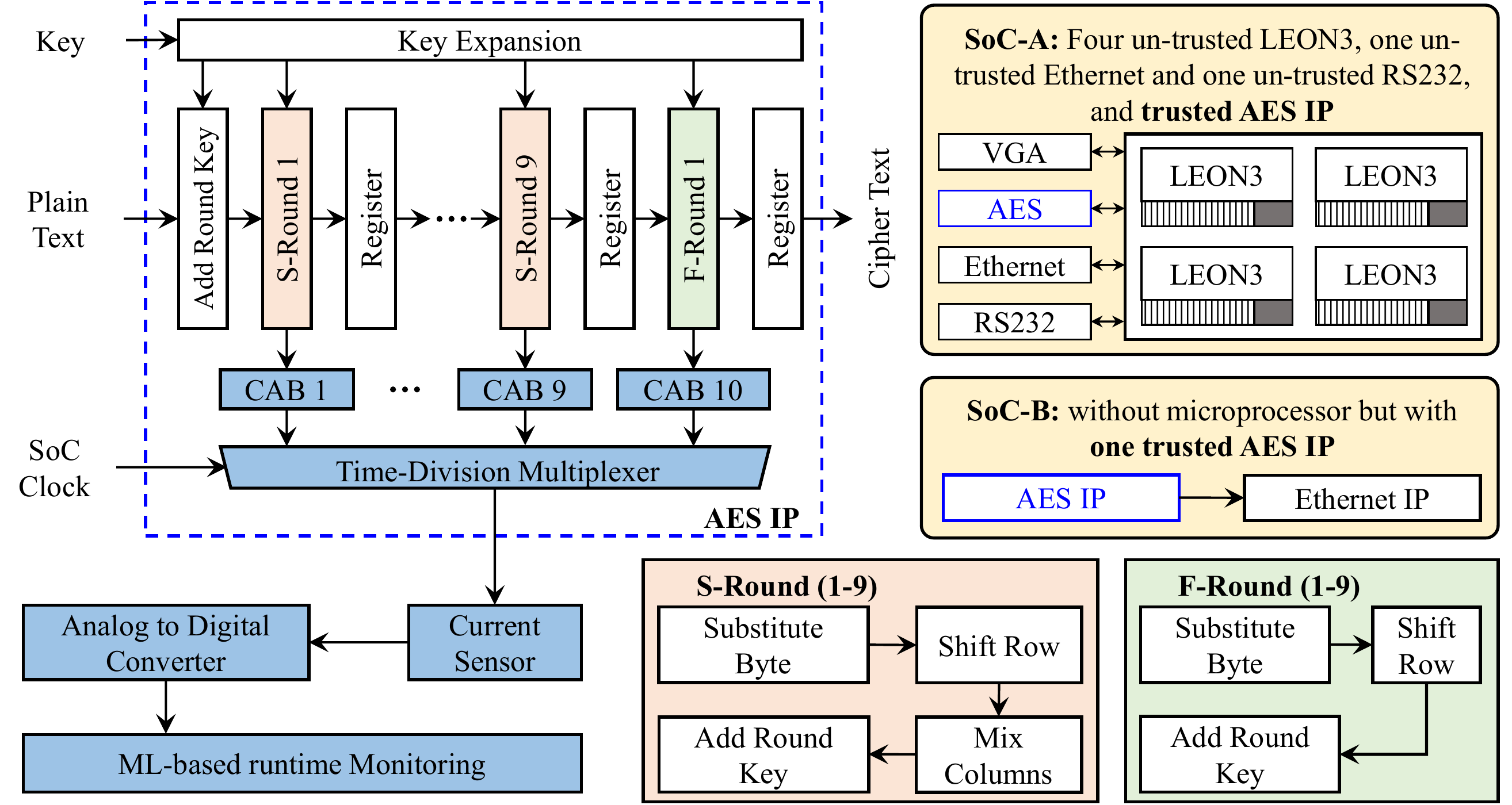} 
	\caption{\textit{SoCs to evaluate the scalability and generalizability of MacLeR. SoC-A consists of 4 un-trusted LEON3 processors , one un-trusted VGA IP, one un-trusted Ethernet IP, one un-trusted RS232 and one trusted AES IP. SoC-B consists of one trusted AES IP and one un-trusted Ethernet IP. SoC-B takes the input, encrypts the input and transmits it via Ethernet.}} 
	\vspace{-1mm}
	\label{fig:SoCs}
\end{figure}
\begin{figure*}[!t]
	\centering 
	\includegraphics[width=1\linewidth]{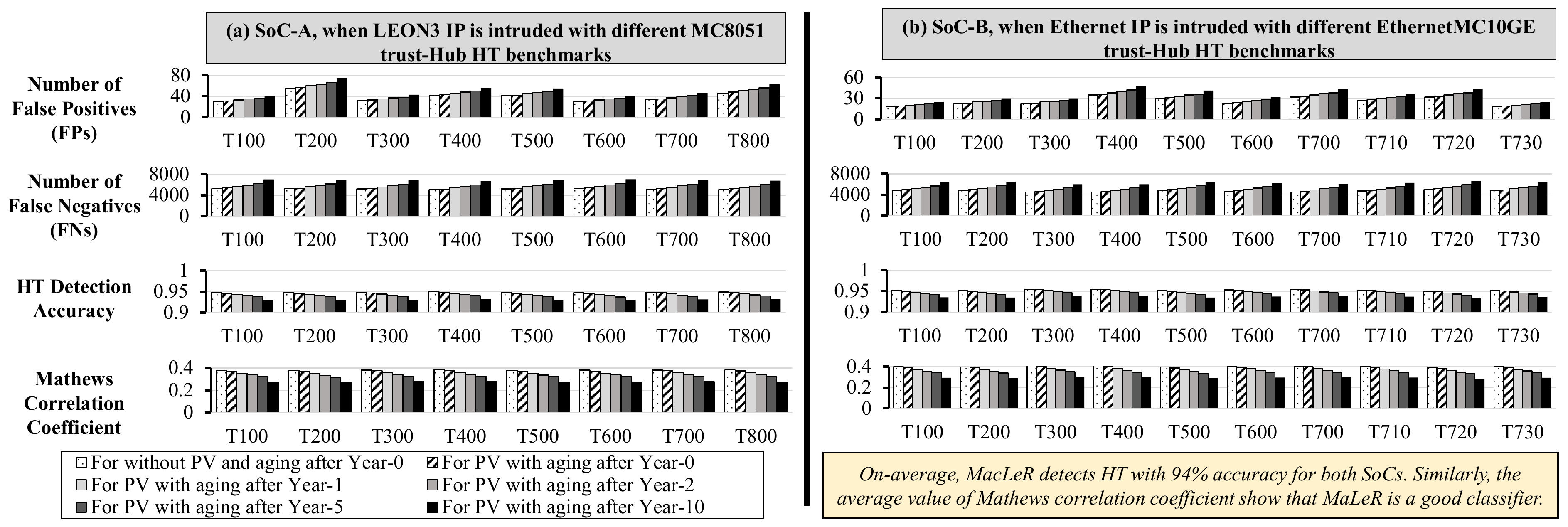} 
	\caption{\textit{Performance analysis McLeR for SoC-A and SoC-B in terms of false positives, false negatives and HT detection accuracy in different scenarios, i.e., without PV and aging, PV without aging, PV with aging after Year-0, 1, 2, 5 and 10. Note, results for SoC-A are for the case when LEON3 is intruded with different MC8051 trust-Hub HT benchmarks, and results for SoC-B are for the case when Ethernet IP is intruded with different HT benchmarks~\cite{trust-HUB}}} 
	\vspace{-3mm}
	\label{fig:SoCs-results}
\end{figure*}
\vspace{-10pt}
\section{Sensitivity Analysis under Aging Effects}\label{sec:aging}
For the comprehensive analysis, we evaluated MacLeR for different HT benchmark under different PV and aging variations based on the model presented in literature~\cite{gnad2015hayat,tiwari2008facelift,rehman2014dtune}, i.e., change in operating frequency after Year-1, Year-2, Year-5, and Year-10. Moreover, we also evaluated MacLeR with no aging mitigation and for two aging policies, i.e., Fast-Core-Age-First and balanced-aging profile.

\begin{enumerate}[leftmargin=*]
    \item Fig~\ref{fig:aging} shows that in the case of no aging mitigation technique, HT detection accuracy drop of MLP trained using MacLeR subjected to PV is 0.44\% and subjected to aging after Year-10 is 8.936\% when compared to the case without any variations. However, the decreasing rate of MCC values due to aging is steeper than HT detection accuracy, which shows that the number of false positives and false negatives increase with aging.
    \item Fig~\ref{fig:aging} shows that in the case of the aging mitigation technique, i.e., Fast-Core-Age-First and balanced-aging profile, the HT detection accuracy drop of MLP trained is relatively smaller as compared to the no-aging policy. Similarly, the decreasing rate of MCC values is significantly lower than the worst-case aging scenario.
\end{enumerate}

\section{Scalability and Generalizability of MacLeR}\label{sec:scalability}
To evaluate the scalability and generalizability of MacLeR, we also evaluated MacLeR on SoC-A (with untrusted LEON3 and trusted AES IP) and SoC-B (non-microprocessor SoC with trusted AES), see Fig,~\ref{fig:SoCs}. Note, in both SoCs, the fine-grained power profiles of trusted AES IP are obtained by computing the power consumption of computation blocks for each round separately, as shown in Fig.~\ref{fig:SoCs}. Experimental results in Fig.~\ref{fig:SoCs-results} show the effectiveness of MacLeR for IP with and without trusted microprocessors. Even in the presence of non-microprocessor trusted IP, MacLeR effectively use the fine-grained power profiles of other non-microprocessor trusted IP, i.e., AES IP, to detect the HT with $\approx$95\% accuracy.

\subsection{Different Workloads running on trusted LEON3 IP}
\begin{figure}[h]
	\centering 
	\includegraphics[width=1\linewidth]{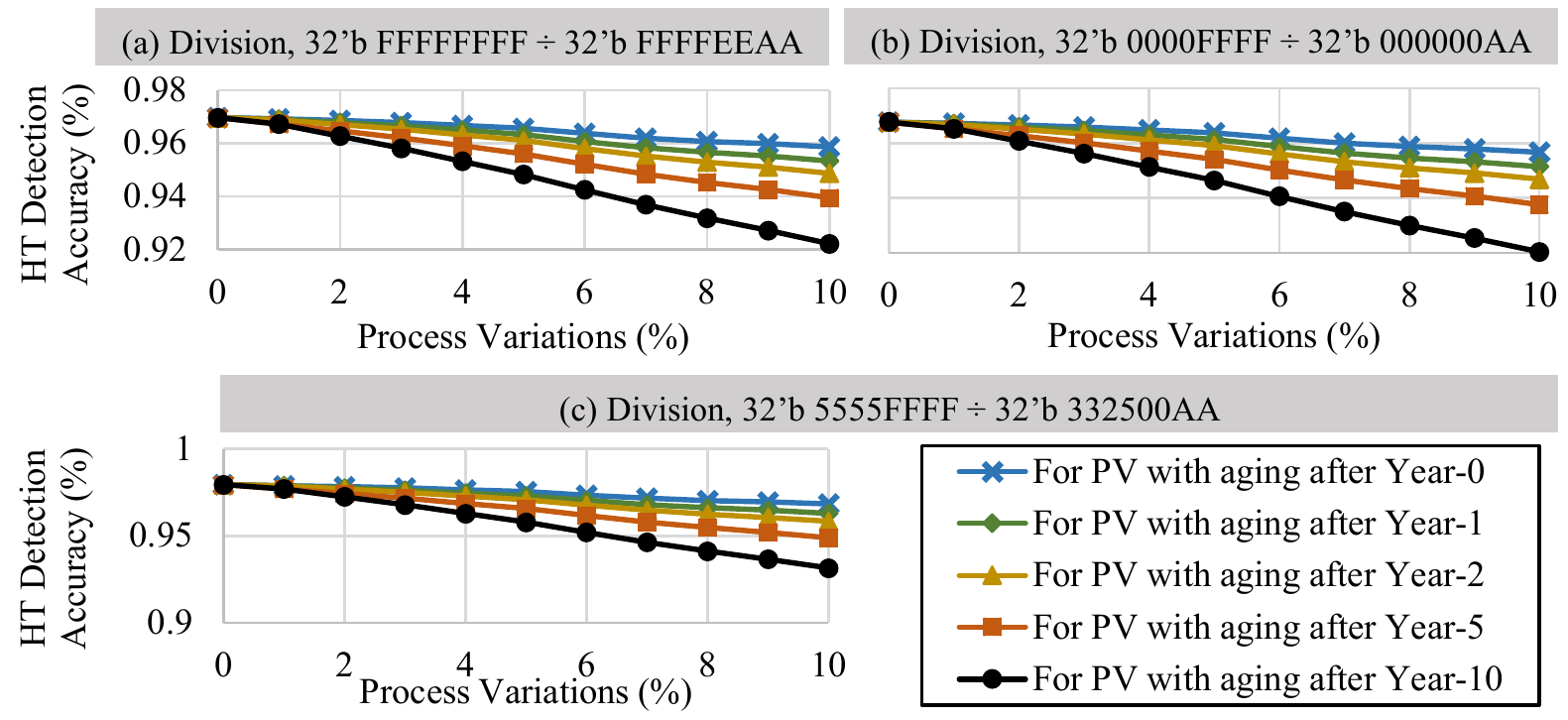} 
	\caption{\textit{HT detection accuracy of MacLeR when different input vectors are running on the trusted LEON3 IP in different scenarios, i.e., without PV and aging, PV without aging, PV with aging after Year-0, 1, 2, 5 and 10.}} 
	\vspace{-1mm}
	\label{fig:LEON3_workload}
\end{figure}

We also evaluated MacLeR, for different input vector to a division workload that is running on a trusted LEON3 IP in LEON3-based SoC. Note, we also considered the PV and different aging effects to evaluate MacLeR for the un-predicted real-world. Fig.~\ref{fig:LEON3_workload} shows the HT detection accuracy when three different input vectors are used for the division workload. The experimental analysis shows that for all input vectors, MacLeR behaves very similar trends, high HT detection accuracy and shows very small variations with respect to input vectors. Hence, MacLeR can effectively detect HTs irrespective of input vectors.
\subsection{Different Workloads running on un-trusted LEON3 IPs}
\begin{figure}[!t]
	\centering 
	\includegraphics[width=1\linewidth]{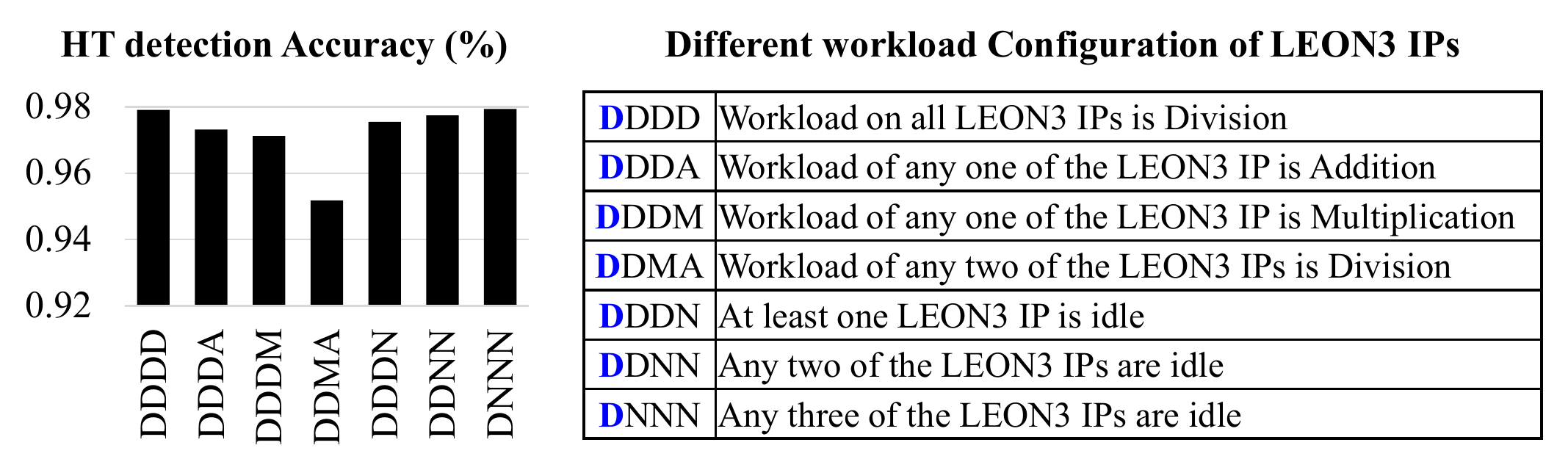} 
	\caption{\textit{HT detection accuracy of MacLeR when different workloads are running on other un-trusted LEON3 IPs. Note, in these experiments, the work load running on trusted LEON3 IP is division.}} 
	\vspace{-1mm}
	\label{fig:workload_config}
\end{figure}

We also evaluated MacLeR for different workload configurations, as shown in Fig.~\ref{fig:workload_config}. In these experiments, the workload in trusted LEON3 IP is fixed to division, and the rest of the un-trusted LEON3 IPs can remain idle, can run multiplication, addition or division, see the different workload configurations in Fig.~\ref{fig:workload_config}. These experimental results show that variation in the workload across the chip have negligible impact (i.e., 2\% to 3\% decrease) on the HT detection accuracy.

\textit{Based on the above-mentioned sensitivity and scalability analyses, we conclude that in most cases, variations in fine-grained power profiles of the microprocessor due to HT are significantly high and distinguishable as compared to the variations due to abrupt changes, PV, aging or fluctuations due to power gating (e.g., blue peaks in Fig.~\ref{fig:PV-2} (b)). Hence these changes are captured by our MacLeR for HT detection. Moreover, depending upon the applications, the frequency and significance of the pipeline stage can be adjusted to increase the HT detection accuracy.}

\section{On-chip and off-chip Overhead of our New Hardware Components of MacLeR} 
To analyze the on-chip overhead of the proposed MacLeR methodology, we synthesized the RTL of the complete LEON3-based SoC using a 65nm technology in Cadence Genus. On the other hand, the power consumption of off-chip components is estimated based on commonly used SoCs, e.g., TMS3280x SoC for ADC. However, the off-chip area overhead is specific to a given platform. Therefore, in this paper, our primary focus is overhead with respect to power consumption. Table~\ref{tab:overhead} provides the on-chip and off-chip area and power overhead for LEON3-based SoC and SoC-A\footnote{Note, this same SoC but the trusted IP is AES instead of LEON3.}. 
    
\textbf{On-chip area overhead:} The overall on-chip area overhead for SP-CAB, time multiplexing in LEON3-based SoC is approximately 70.23$\mu m^2$, and the on-chip area overhead of SoC-A is approximately 103.26$\mu m^2$. The reason behind the extra overhead for the latter is that there are more SP-CABs in SoC-A as compared to SP-CABs in LEON3-based IP. In summary, the area overheads for LEON3-based SoC and SoC-A \textit{is less than 0.025\% of the total area, thus negligible.} This analysis also shows that area overheads of the trusted IPs, i.e., LEON3 in LEON3-based SoC and AES in SoC-A, is also 0.06\% in LEON3 IP and 0.5\% in AES, thus, negligible.
    
\textbf{On-chip and Off-chip power overhead:} Similarly, the on-chip power overhead for LEON3-based SoC is $\approx$0.47 $mW$, and the on-chip power overhead of SoC-A is 0.57 $mW$. \textit{which is less than 1\% of the total power, thus negligible}. In summary, the on-chip power overheads for LEON3-based SoC and SoC-A \textit{is less than 0.07\% of the total area, thus negligible.} However, the off-chip power overheads for LEON3-based SoC and SoC-A are 36.74 $mW$ and 30.26 $mW$, respectively. The overall on-chip and off-chip power overhead is less than 5\% of the total power consumption; thus, it can be considered as tolerable ~\cite{xiao2016hardware}. Similarly, the power overheads of the trusted IPs, i.e., LEON3 in LEON3-based SoC and AES in SoC-A, are also 0.06\% in LEON3 IP and 0.5\% in AES, respectively, thus negligible. \normalcolor \textbf{Summarizing the key benefits:}
\begin{enumerate}[leftmargin=*]
    \item The area overhead of MacLeR is 7x less in terms of power ports as compared to state-of-the-art ~\cite{lodhi2017power}\cite{bao2014application}\cite{bao2016reverse}\cite{he2017hardware}. Moreover, the best-selected MLP configuration for MacLeR consists of two hidden layers, and each layer has only 8 neurons. On the other hand, the MLP in~\cite{lodhi2017power} with maximum HT detection accuracy (i.e., 85.12\%) consists of 50 hidden layers, and each layer has 100 neurons. 

    \item The overhead in terms of physical area is larger than the technique of \cite{bao2014application}\cite{bao2016reverse}. However, this technique uses reverse engineering, which increases the complexity, time, effort, and cost substantially. Our technique, in contrast, uses a very simple design and a minimal area overhead, making it practical for fast deployment for the IoT edge devices. 
   
\end{enumerate}

\begin{table}[!t]
        \caption{\textit{On-chip and off-chip area and power overhead analysis}}
    	\label{tab:overhead}
    	\resizebox{1\linewidth}{!}{
        \begin{tabular}{lllll}
            \cline{2-5}
            \multicolumn{1}{l|}{} & \multicolumn{4}{c|}{\textbf{Area ($\mu m^2$)}} \\ \cline{2-5} 
            \multicolumn{1}{l|}{} & \multicolumn{1}{l|}{LEON3-based SoC} & \multicolumn{1}{l|}{SoC-A} & \multicolumn{1}{l|}{LEON3 IP} & \multicolumn{1}{l|}{AES IP} \\ \hline
            \multicolumn{1}{|l|}{\textbf{Without overhead}} & \multicolumn{1}{l|}{464851.302} & \multicolumn{1}{l|}{464851.302} & \multicolumn{1}{l|}{108330.348} & \multicolumn{1}{l|}{20332.8} \\ \hline
            \multicolumn{1}{|l|}{\textbf{With on-chip overhead}} & \multicolumn{1}{l|}{464921.532} & \multicolumn{1}{l|}{464954.562} & \multicolumn{1}{l|}{108400.578} & \multicolumn{1}{l|}{20436.06} \\ \hline
            \multicolumn{1}{|l|}{\textbf{With on-chip \& off-chip overhead}} & \multicolumn{1}{l|}{N/A} & \multicolumn{1}{l|}{N/A} & \multicolumn{1}{l|}{N/A} & \multicolumn{1}{l|}{N/A} \\ \hline
             &  &  &  &  \\ \cline{2-5} 
            \multicolumn{1}{l|}{} & \multicolumn{4}{c|}{\textbf{Power ($mW$)}} \\ \cline{2-5} 
            \multicolumn{1}{l|}{} & \multicolumn{1}{l|}{LEON3-based SoC} & \multicolumn{1}{l|}{SoC-A} & \multicolumn{1}{l|}{LEON3 IP} & \multicolumn{1}{l|}{AES IP} \\ \hline
            \multicolumn{1}{|l|}{\textbf{Without overhead}} & \multicolumn{1}{l|}{734.180657} & \multicolumn{1}{l|}{734.180657} & \multicolumn{1}{l|}{183.2784452} & \multicolumn{1}{l|}{0.688} \\ \hline
            \multicolumn{1}{|l|}{\textbf{With on-chip overhead}} & \multicolumn{1}{l|}{734.6559309} & \multicolumn{1}{l|}{734.755157} & \multicolumn{1}{l|}{183.3972637} & \multicolumn{1}{l|}{0.6915} \\ \hline
            \multicolumn{1}{|l|}{\textbf{With on-chip \& off-chip overhead}} & \multicolumn{1}{l|}{770.9199309} & \multicolumn{1}{l|}{764.444157} & \multicolumn{1}{l|}{N/A} & \multicolumn{1}{l|}{N/A} \\ \hline
        \end{tabular}}
    \end{table}
\section{Conclusion}
\label{conclusion}
This paper presents a methodology to design an ML-based run-time HTs detection (MacLeR) for resource-constrained IoT edge devices. MacLeR first extracts the instruction-dependent fined-grained power profile of different pipeline stages. Afterwards, it addresses the challenging problem of run-time data acquisition by designing a single power-port based current acquisition block. The power profiles are used to train and explore the design space of multiple ML models, and to select the model providing the highest HT detection accuracy. To illustrate the scalability and generalizability of the MacLeR, we evaluated it on different SoCs with microprocessors as trusted IP, non-microprocessor trusted IP, and non-microprocessor SoCs. Our experimental results show that as compared to the state-of-the-art HT detection technique, MacLeR achieves a 10\% increase in HT detection accuracy (i.e., 96.256\%), while incurring 7x reduction area and power overhead. We also analyzed the impact of process variations and aging variations on MacLeR. The analysis shows that With proper aging policies and PV consideration can effectively, MacLeR can handle the unpredictability of real-world applications. Hence, this simple design with negligible area/power overhead and high tolerance against process variation makes MacLeR feasible in real-world IoT-edge and CPS devices. 
\section*{Acknowledgment}
This work is supported in parts by the Austrian Research Promotion Agency (FFG) and the Austrian Federal Ministry for Transport, Innovation, and Technology (BMVIT) under the ``ICT of the Future'' project, IoT4CPS: Trustworthy IoT for Cyber-Physical Systems.
%=====================================================================
%\small
\bibliographystyle{IEEEtran}
\bibliography{MacLeR.bbl}
\vspace{-0.4in}
\begin{IEEEbiography}[{\includegraphics[width=1in,height=1.25in,clip,keepaspectratio]{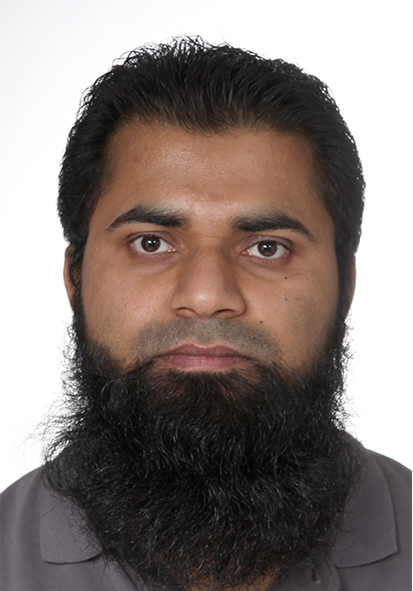}}]{Faiq Khalid}(S'18) received his M.S. degree in electrical engineering and his B.E. degree in electronics engineering from the National University of Sciences and Technology (NUST), Pakistan, in 2016 and in 2011, respectively. He is currently pursuing his Ph.D. degree in hardware security and machine learning security at Technische Universit{\"a}t Wien (TU Wien), Vienna, Austria. He is a recipient of the Quaid-e-Azam Gold Medal for his academic achievements, the Best Researcher Award at the SAVe Lab in 2014, and the Richard Newton Young Fellowship Award at DAC 2018. His research interests include formal analysis and verification of embedded systems, hardware design security, and security for machine learning systems. He has served as a TPC member of several FIT, WSAV and ARES, and a regular reviewer of several prestigious journals (e.g., IEEE Access, TCAD, TCAS-I, TCAS-II, TVLSI) and conferences. 
\end{IEEEbiography}
\vskip -2\baselineskip plus -1fil
\begin{IEEEbiography}[{\includegraphics[width=1in,height=1.25in,clip,keepaspectratio]{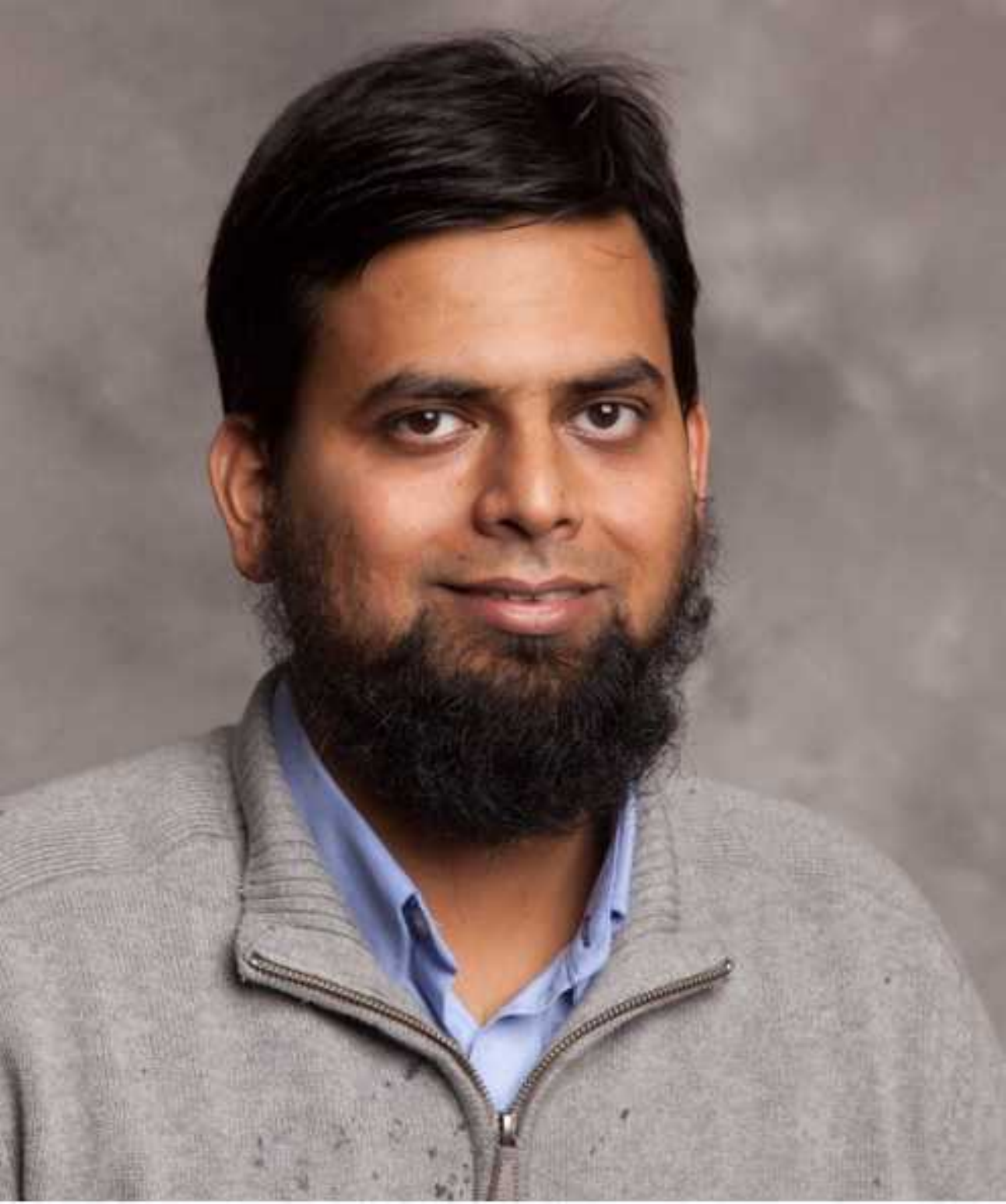}}]{Syed Rafay Hasan} received the B.Eng. Degree in electrical engineering from the NED University of Engineering and Technology, Pakistan, and the M.Eng. and Ph.D. degrees in electrical engineering from Concordia University, Montreal, QC, Canada. He served as an Adjunct Faculty Member with Concordia University (2006-2009) and a Research Associate with the Ecole Polytechnique de Montreal (2009-2011). Since 2011, he has been with the Electrical and Computer Engineering Department, Tennessee Tech University, Cookeville, TN, USA, where he is currently an Associate Professor. He has published 70+ peer-reviewed journal and conference papers. His current research interests include hardware design security in the Internet of Things (IoT), hardware implementation of deep learning, and edge intelligence, including the deployment of CNN in the IoT edge devices. He received the Postdoctoral Fellowship Award from the Scholarship Regroupment Stratgique en Microsystmes du Québec, the Faculty Research Award, Sigma Xi Outstanding Research Award from Tennessee Tech University, the Kinslow Outstanding Research Paper Award from the College of Engineering, Tennessee Tech University, and the Summer Faculty Fellowship Award from the Air force Research Lab (AFRL). He has received research and teaching funding from NSF, ICT-funds UAE, AFRL, and Intel Inc. He has been part of the funded research projects, as a PI or a Co-PI, that worth more than \$1.2 million. He has been the on NSF panel of reviewers, session chair or technical program committee member of several IEEE conferences including ISCAS, ICCD, MWSCAS, and NEWCAS, and a regular reviewer IEEE Transactions and other journals including TCAS-II, IEEE ACCESS, IEEE Embedded Systems Letters, Integration, the VLSI Journal, IET Circuit Devices and Systems, and IEEE Transaction on Neural Networks. He is a Full Member of Sigma Xi and a Life Member of the Pakistan Engineering Council. 
\end{IEEEbiography}
%\vskip -2\baselineskip plus -1fil
\begin{IEEEbiography}[{\includegraphics[width=1in,height=1.25in,clip,keepaspectratio]{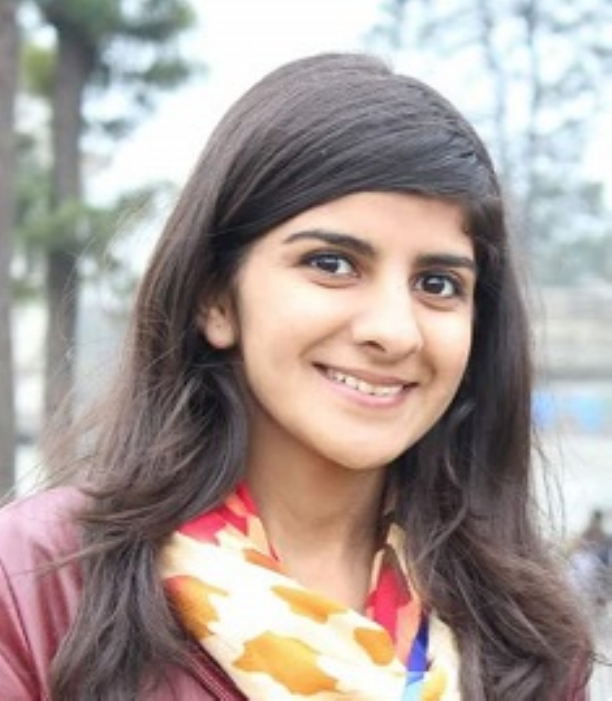}}]{Sara Zia} received the B.E. and M.S. Electrical Engineering degree from National University of Sciences and Technology (NUST), Pakistan in 2015 and 2018 respectively. She worked as a Research Assistant in System Analysis and Verification (SAVe) Lab at NUST School of Electrical Engineering and Computer Sciences. Her research interests include digital systems and hardware design security. She is a member of Pakistan Engineering Council.
\end{IEEEbiography}
\vskip -2\baselineskip plus -1fil
\begin{IEEEbiography}[{\includegraphics[width=1in,height=1.2in,clip]{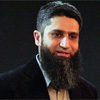}}]{Osman Hasan} received the M.Eng. and Ph.D. degrees from Concordia University, Montreal, Canada, in 2001 and 2008, respectively. Currently, he is an Associate Professor at the National University of Science and Technology (NUST), Islamabad, Pakistan and the Principal and Dean of NUST-SEECs. He is the founder and director of System Analysis and Verification (SAVe) Lab at NUST, which mainly focuses on the design and formal verification of safety-critical systems. He has received several awards and distinctions, including the Pakistan’s Higher Education Commission’s Best University Teacher (2010) and Best Young Researcher Award (2011) and the President’s gold medal for the best teacher of the University from NUST in 2015. He is a senior member of IEEE, member of the ACM, AAR and the Pakistan Engineering Council.
\end{IEEEbiography}
\vskip -2\baselineskip plus -1fil
\begin{IEEEbiography}[{\includegraphics[width=1in,height=1.25in,clip,keepaspectratio]{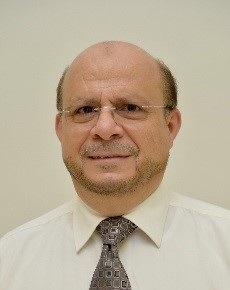}}]{Falah Awwad} received the MSc. and Ph.D. degrees in Electrical and Computer Engineering from Concordia University (Montreal, QC, Canada) in 2002 and 2006, respectively. He was a Post-Doctoral Fellow at Ecole Polytechnique de Montréal and Concordia University, Montreal, QC, Canada. Between August 2007 and Feb. 2013, he was an Assistant Professor with the College of Information Technology at United Arab Emirates University. Currently, he is a Professor with the Department of Electrical Engineering - College of Engineering (UAE University). He published 80+ papers in premier journals and conferences. He is the PI and Co-PI of 20 research projects and supervised several postgraduate students.  He served as the Session Chair and a TPC member of several prestigious conferences. He is a member of the editorial Boards and regular reviewer of several prestigious Journals. His scientific research interests include primarily sensors, circuits, and devices, in addition to hardware security, biomedical applications, and smart grids.
\end{IEEEbiography}
\vskip -2\baselineskip plus -1fil
\begin{IEEEbiography}[{\includegraphics[width=1in,height=1.25in,clip,keepaspectratio]{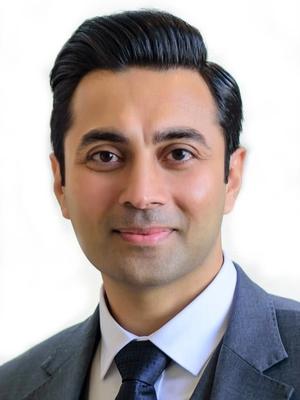}}]
{Muhammad Shafique}(M'11 - SM'16) received the Ph.D. degree in computer science from the Karlsruhe Institute of Technology (KIT), Germany, in 2011. Afterwards, he established and led a highly recognized research group at KIT for several years as well as conducted impactful R\&D activities in Pakistan. In Oct.2016, he joined the Institute of Computer Engineering at the Faculty of Informatics, Technische Universität Wien (TU Wien), Vienna, Austria as a Full Professor of Computer Architecture and Robust, Energy-Efficient Technologies. Since Sep.2020, he is with the Division of Engineering, New York University Abu Dhabi (NYU AD), United Arab Emirates.

His research interests are in brain-inspired computing, AI \& machine learning hardware and system-level design, energy-efficient systems, robust computing, hardware security, emerging technologies, FPGAs, MPSoCs, and embedded systems. His research has a special focus on cross-layer analysis, modeling, design, and optimization of computing and memory systems. The researched technologies and tools are deployed in application use cases from Internet-of-Things (IoT), smart Cyber-Physical Systems (CPS), and ICT for Development (ICT4D) domains.

Dr. Shafique has given several Keynotes, Invited Talks, and Tutorials, as well as organized many special sessions at premier venues. He has served as the PC Chair, Track Chair, and PC member for several prestigious IEEE/ACM conferences. 
Dr. Shafique holds one U.S. patent has (co-)authored 6 Books, 10+ Book Chapters, and over 200 papers in premier journals and conferences. He received the 2015 ACM/SIGDA Outstanding New Faculty Award, AI 2000 Chip Technology Most Influential Scholar Award in 2020, six gold medals, and several best paper awards and nominations at prestigious conferences. \end{IEEEbiography}
%=====================================================================

\end{document}